%% file: 0_main_tosem22.tex
\documentclass[acmsmall, authorversion, nonacm]{acmart}

\AtBeginDocument{%
  \providecommand\BibTeX{{%
    \normalfont B\kern-0.5em{\scshape i\kern-0.25em b}\kern-0.8em\TeX}}}

\setcopyright{acmlicensed}
\acmJournal{TOSEM}
\input{latex_commands/config}

\renewenvironment{updated}{\par}{\par}
\renewcommand{\updatedtext}[1]{#1}


\newcommand\blfootnote[1]{%
  \begingroup
  \renewcommand\thefootnote{}\footnote{#1}%
  \addtocounter{footnote}{-1}%
  \endgroup
}

\begin{document}

\title{Programming by Example Made Easy}

  \author{Jiarong Wu}
  \affiliation{%
    \institution{Department of Computer Science and Engineering, The Hong Kong University of Science and Technology}
    \city{Hong Kong}
    \country{China}}
  \email{jwubf@cse.ust.hk}

  \author{Lili Wei}
  \affiliation{%
    \institution{Department of Electrical and Computer Engineering, McGill University}
    \city{Montreal}
    \country{Canada}}
  \email{lili.wei@mcgill.ca}

  \author{Yanyan Jiang}
  \authornote{Corresponding authors}
  \affiliation{%
    \institution{State Key Laboratory for Novel Software Technology and Department of Computer Science and Technology, Nanjing University}
    \city{Nanjing}
    \country{China}}
  \email{jyy@nju.edu.cn}

  \author{Shing-Chi Cheung}
  \authornotemark[1]
  \affiliation{%
    \institution{Department of Computer Science and Engineering, The Hong Kong University of Science and Technology}
    \city{Hong Kong}
    \country{China}}
  \email{scc@cse.ust.hk}

  \author{Luyao Ren}
  \affiliation{%
    \institution{Key Lab of High Confidence Software Technologies, Ministry of Education Department of Computer Science and Technology, EECS, Peking University}
    \city{Beijing}
    \country{China}}
  \email{rly@pku.edu.cn}

  \author{Chang Xu}
  \affiliation{%
    \institution{State Key Laboratory for Novel Software Technology and Department of Computer Science and Technology, Nanjing University}
    \city{Nanjing}
    \country{China}}
  \email{changxu@nju.edu.cn}

  \begin{abstract}
    Programming by example (PBE) is an emerging programming paradigm that automatically synthesizes programs specified by user-provided input-output examples.
    Despite the convenience for end-users,
    implementing PBE tools often requires strong expertise in programming language and synthesis algorithms.
    Such a level of knowledge is uncommon among software developers.
    It greatly limits the broad adoption of PBE by the industry.
    To facilitate the adoption of PBE techniques, we propose a PBE framework called \proj,
    which leverages an ``entity-action'' model based on relational tables to ease PBE development for a wide but \updatedtext{restrained} range of domains.
    Implementing PBE tools with \proj only requires adapting domain-specific data entities and user actions to tables,
    with no need to design a domain-specific language or an efficient synthesis algorithm.
    The synthesis algorithm of \proj exploits bidirectional searching and constraint-solving techniques to address the challenge of value computation nested in table transformation.
    We evaluated \proj's effectiveness on 64 PBE tasks from three different domains and usability with a human study of 12 participants.
    Evaluation results show that \proj is easier to learn and use than the state-of-the-art PBE framework,
    and the bidirectional algorithm achieves comparable performance to domain-specifically optimized synthesizers.

    \blfootnote{This paper has been accepted by ACM Transactions on Software Engineering and Methodology.}
  \end{abstract}

  \maketitle

  \input{latex_commands/symbols}
  \input{1_introduction}
  \input{2_overview}
  \input{3_specification}
  \input{4_algorithm}

  \input{5_evaluation}

  \input{6_ending}
  \input{7_ack}

  \bibliography{pbe}

  \appendix
  \input{a1_example}
  \input{a2_evaluating_specs}

\end{document}

%% file: latex_commands/config.tex
\usepackage{multirow}
\usepackage{subcaption}
\usepackage{xspace}
\usepackage{color, soul, xcolor}
\usepackage{pdfpages}
\usepackage[vlined,ruled,linesnumbered]{algorithm2e}
\usepackage[noend]{algpseudocode}
\usepackage{cleveref}
\usepackage{amsmath}
\usepackage{mathtools}
\usepackage{ifthen}
\usepackage{adjustbox}
\usepackage{booktabs}
\usepackage{threeparttable}
\usepackage{url}
\usepackage{etoolbox}
\usepackage{pifont}
\newtheorem{theorem}{Theorem}

\definecolor{darkgreen}{rgb}{0,0.4,0}

\newcommand{\etc}{{etc}}
\newcommand{\eg}{{e.g.}}
\newcommand{\ie}{{i.e.}}

\newcommand{\proj}{\textsc{Bee}\xspace}
\newcommand{\projf}{\proj-f\xspace}
\newcommand{\prose}{\textsc{Prose}\xspace}
\newcommand{\trinity}{\textsc{Trinity}\xspace}
\newcommand{\hades}{\textsc{Hades}\xspace}
\newcommand{\rosette}{\textsc{Rosette}\xspace}

\newcommand{\synthfunc}{\textsc{Bee$_S$}\xspace}
\newcommand{\lang}{\textsc{Bee$_L$}\xspace}
\newcommand{\duet}{\textsc{Duet}\xspace}
\newcommand{\viser}{\textsc{Viser}\xspace}

\newcommand{\hasComment}[0]{true}
\ifthenelse{\equal{\hasComment}{true}}
{%
    \newcommand{\todoc}[2]{\textcolor{#1}{[#2]}}
}
{%
    \newcommand{\todoc}[2]{}
}

\newenvironment{updated}{\par\color{purple}}{\par}
\newcommand{\updatedtext}[1]{\textcolor{purple}{#1}}

\usepackage{listings}
\definecolor{codeblue}{rgb}{0,0,0.9}
\definecolor{codegray}{rgb}{0.5,0.5,0.5}
\definecolor{codepurple}{rgb}{0.58,0,0.82}
\definecolor{backcolour}{rgb}{0.95,0.95,0.92}

\lstdefinestyle{mystyle}{
    basicstyle=\scriptsize\ttfamily,
    backgroundcolor=\color{white},
    commentstyle=\color{codegray},
    keywordstyle=\color{magenta},
    numberstyle=\tiny\color{codegray},
    stringstyle=\color{codepurple},
    breakatwhitespace=false,
    breaklines=false,
    captionpos=b,
    keepspaces=true,
    numbers=left,
    numbersep=10pt,
    showspaces=false,
    showstringspaces=false,
    showtabs=false,
    tabsize=2,
    frame=single
}
\usepackage{enumitem}

\crefname{lstlisting}{listing}{listings}
\Crefname{lstlisting}{Listing}{Listings}

\newcommand{\smalltitle}[1]{\smallskip\noindent\textbf{#1.}}
\newcommand{\construct}[1]{\smallskip\noindent\textit{#1}} %
\newcommand{\codef}[1]{\texttt{#1}}

\newcounter{reviewer}
\newcounter{point}[reviewer]
\newcommand\shorttitle[0]{}

\renewcommand{\thepoint}{\thereviewer.\arabic{point}}

\newcommand{\shortreply}[2][]{\medskip \noindent \begin{sf} \textbf{Reply}:\  #2
\ifthenelse{\equal{#1}{}}{}{ \hfill \footnotesize (#1)}%
\medskip \end{sf}}

%% file: latex_commands/symbols.tex
\newcommand{\cmark}{\ding{51}}%
\newcommand{\xmark}{\ding{55}}%

\newcommand{\texi}{T^i}
\newcommand{\texo}{t^o}
\newcommand{\tpend}{T^p}
\newcommand{\ttarget}{t^s}
\newcommand{\constpool}{C}

\newcommand{\tuple}[1]{\langle #1 \rangle}
\newcommand{\type}[1]{\texttt{#1}}
\newcommand{\typeid}{\type{Id}\xspace}
\newcommand{\typestr}{\type{String}\xspace}
\newcommand{\typeint}{\type{Int}\xspace}

\newcommand{\symname}{\eta}
\newcommand{\Tau}{\mathrm{T}}
\newcommand{\symschema}{\Tau}
\newcommand{\symschemao}{\mathrm{T}^o}

\newcommand{\append}{\mathbin{+\mkern-10mu+}}

\newcommand{\row}{\mathit{row}}
\newcommand{\col}{\mathit{col}}

\newcommand{\symlang}{L}

\newcommand{\symprog}{P}
\newcommand{\sympt}{P_t}
\newcommand{\sympm}{P_m}
\newcommand{\eval}[1]{\llbracket #1 \rrbracket}

\newcommand{\altab}[0]{$\ |\ $\xspace}
\newcommand{\symst}{s_t}
\newcommand{\symsm}{s_m}
\newcommand{\pred}{\phi}
\newcommand{\aggr}{\alpha}
\newcommand{\projexpr}{\rho}
\newcommand{\mutate}{\mu}
\newcommand{\pj}{\rho} %
\newcommand{\highfeature}{f_h}
\newcommand{\opFilter}{Filter\xspace}
\newcommand{\opJoin}{Join\xspace}
\newcommand{\opGroup}{GroupJoin\xspace}
\newcommand{\opOrder}{Order\xspace}
\newcommand{\opYield}{Yield\xspace}
\newcommand{\opMutate}{Mutate\xspace}

\newcommand{\tpja}{t^\pj_1}
\newcommand{\tpjb}{t^\pj_2}
\newcommand{\tout}{t^{out}}

\newcommand{\symstate}{\Sigma}

\newcommand{\ptset}{\overline{\symst}}
\newcommand{\pmset}{\overline{\symsm}}
\newcommand{\hole}{\square}
\newcommand{\encode}[1]{\varphi(#1)}
\newcommand{\paramset}{\theta}

\newcommand{\actionShift}[4]{$\langle$shift, \codef{#1}, #2, $#3$, $#4\rangle$\xspace}

%% file: 1_introduction.tex
\section{Introduction}\label{sec:introduction}

\smalltitle{Background and Motivation}
Given a set of program inputs and their expected corresponding outputs as \emph{examples},
programming by example (PBE) is a program synthesis paradigm that can automatically deduce a program conforming to the provided examples~\cite{halbert1984, synthesis-survey}.
PBE has been widely demonstrated to be successful in automating domain-specific repetitive end-user tasks~\cite{flashfill11, visual19, extract14, extract16, extract17, edit19, cad20, merge_conflict21}. %
Despite such reported successes, PBE is seldom a supported feature in mainstream software products.
Thus far, each PBE implementation requires a concise domain-specific language for the target application and an algorithm optimized for the language.
The skills involved are out of reach by many developers, hindering the wide adoption of PBE\@.

\begin{figure}
    \begin{center}
        \includegraphics[page=1, width=\linewidth]{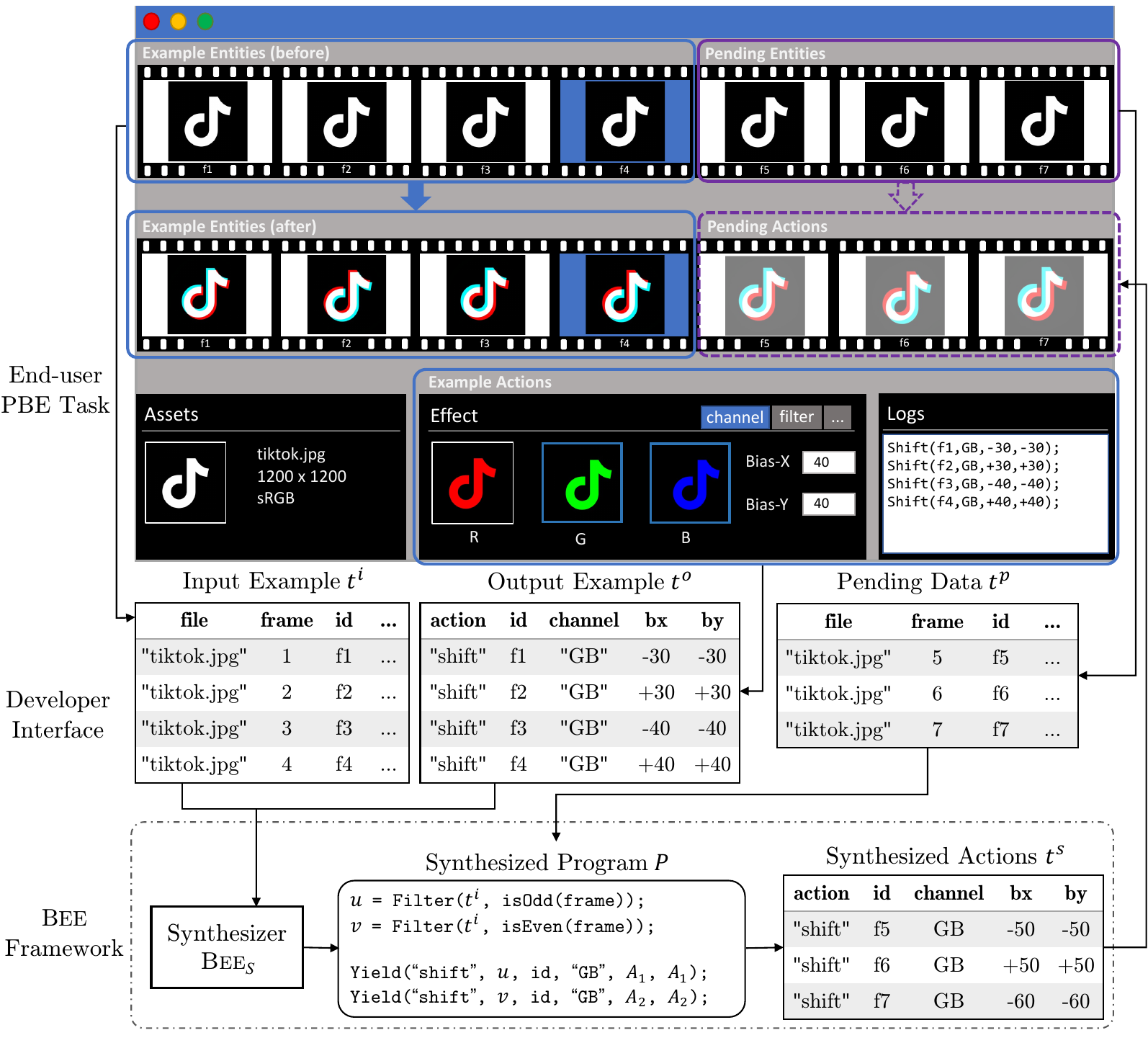}
    \end{center}
    \caption{A PBE task of GIF editing in a hypothetical video editing application, in which a designer creates a TikTok-style GIF animation~\cite{beehomepage} by providing a few example frame editing actions.
    With the table representations of example frames/actions modeled by the application developer,
    \proj generalizes the examples and synthesizes a program $\symprog$ (semantics available in \Cref{fig:execution}) to automate the conversion of remaining frames.
    The GUI is fabricated, but the user actions are adapted from ImageMagick~\cite{magick} commands.}
    \label{fig:task-io}
\end{figure}

Let us illustrate the capability of PBE and the skills involved in supporting such a capability using a GIF editing task.
Suppose that Cathy wants to create an animated GIF in TikTok style as shown in \Cref{fig:task-io}.
Existing video editing software allows Cathy to place copies of an image (\texttt{tiktok.jpg}) on a timeline and apply graphical effects to the image copies.
However, when the effects follow a non-trivial pattern, in which the green-blue channel shifts in opposite directions for odd/even frames with increasing amplitude,
Cathy has to manually apply the effects \emph{separately for each image copy}, which is labor-intensive and error-prone.
Existing industrial video editing software provides no easy way to create such effects.

The editing task can be automated using PBE. %
With appropriate customization,
existing PBE techniques are capable of synthesizing programs to automate effects creation for each frame.
Below is an example code snippet in C\# style.

\begin{lstlisting}[language={[Sharp]C}]
int b = 20 + (frame_id + 1) / 2 * 10;  // calculating the biases in pixels
return (frame_id % 2) switch {
    0 => shift("GB", b, b),    // for even frames, shift down and right (positive biases)
    1 => shift("GB", -b, -b)   // for odd frames, shift up and left (negative biases)
}
\end{lstlisting}

In general, software developers have three options with which to incorporate PBE support (like the example above) into their tools.
\begin{enumerate}
\item Implement the PBE feature from scratch, which is a typical approach in the research literature~\cite{flashfill11, visual19, file16, cad20}.
This requires designing a domain-specific language (DSL) and an algorithm to synthesize programs satisfying the user-given examples~\cite{sygus}.
\item Adopt a PBE framework (\eg, \prose~\cite{flashmeta15, flashmeta18neural} or \trinity~\cite{trinity}) by designing a DSL and customizing the framework's built-in synthesis algorithm.
\item Reuse existing PBE tools whose DSLs offer the desired expressiveness. For example, the code snippet above requires a PBE tool supporting conditional integral arithmetics.
\end{enumerate}

However, these options require being acquainted with DSL and algorithm design.
Such a requirement is not commonly amenable to software developers.  %
Existing research~\cite{flashmeta15} indicates that implementing a PBE feature is time-consuming, even for PBE experts.%
\footnote{
This is confirmed by our human study results (\Cref{subsec:user-study}), where participants, who are experienced developers but without PBE expertise, found the state-of-the-art PBE framework, \prose, difficult to learn and use. }
Moreover, existing PBE implementations are usually research artifacts specifically built for a specific domain.
Adapting them beyond these specific domains is non-trivial.

We propose to bridge the gap with a novel PBE framework, \proj, which
enables software developers to quickly implement efficient PBE tools via an easy-to-use interface without expertise in DSL and algorithm design.
\proj can be applied to PBE tasks in a variety of domains and scenarios.

\smalltitle{Programming by Example Made Easy}
As discussed above, a major obstacle to PBE implementation is its requirement to design a concise DSL and a synthesis algorithm optimized for the DSL.
The DSL is so designed to concisely express domain-specific tasks and allow for a tractable synthesis procedure. %

\proj relieves the effort of DSL design with a predefined SQL-like meta DSL for a class of PBE tasks that \emph{perform repetitive user actions over a large number of similar data entities}.
This class is bounded but is also inclusive and meets the mainstream application scenario of PBE.
Software developers using \proj %
do not need to modify/customize this meta DSL. To ease presentation, we refer to these developers using PBE frameworks to implement the PBE features as \textit{PBE developers}. %
The adoption of a SQL-like meta DSL for \proj is inspired by the observation that scripts for automating repetitive actions over similar entities usually comprise two parts:

\begin{enumerate}
\item Representation of data entities and user actions (\eg, frames and channel shifts in \Cref{fig:task-io}).
\emph{Relational tables} (or simply tables) containing cells of primitive types (\eg, strings and integers) are suitable for uniformly representing data entities,
as evidenced by the success of the entity-relation model (relational databases and ORM~\cite{orm}) in a wide variety of application domains.
Actions performed over data entities can also be represented in the form of tables, where the first column in a table specifies an action type and the remaining columns specify action arguments.

\item Constructs for program logic description.
Most of the logic behind applying repetitive actions over a set of entities with the same data structure can be mechanically expressed as below.
(a) Transform the data entities (tables) into views (tables), each of which holds the data entities that are subject to the same kind of user actions, e.g., two views separately holding odd and even frames for the task in \Cref{fig:task-io}.
Such transformations can be achieved by SQL-like operators (\eg, \texttt{WHERE} and \texttt{JOIN}).
(b) The desired actions (recall that actions are represented as tables) are projected from the views like \texttt{SELECT} in SQL.
In addition to column-to-column projection, \proj allows the generation of a column (action argument) by
applying a computation operator (\eg, linear arithmetic) on certain columns, \eg, bias to shift in \Cref{fig:task-io} is computed on column \codef{frame}.
Specifically, \proj limits the computation in a predefined set of commonly used string and integer operators for synthesis tractability.
\end{enumerate}

With \proj shipping the meta DSL for program logic, PBE developers can integrate a PBE feature (like ``intelligent task automation'') into their application by modeling domain-specific data entities and user actions as relational tables.
Take the PBE task in \Cref{fig:task-io} for example, PBE developers can:
(1) model the example input frames as tables with self-designed schemas that include the key properties needed in the task (\eg, \codef{frame})\footnote{Further guidelines for schema design are available in \proj's tutorial~\cite{beehomepage}.} and
(2) model the example output user actions as a table, e.g.,
shifting the green and blue channels of a frame \codef{f1} 30 pixels up and left can be modeled as a row \actionShift{f1}{GB}{-30}{-30}.

Given the input-output examples as tables (\proj provides ORM-style APIs~\cite{orm} for easy Python/C\# integration), \proj takes care of the program synthesis by producing a \emph{consistent} program that executes on the example input table and yields the example output table.
The data entities pending to be processed would be represented as tables with the same schemas of example inputs.
Then, the tables of pending-data can be fed to the synthesized program to generate the desired user actions.
Specifically, the generated user actions are represented as a table with the same schema of example output.
The software can follow each row in the generated user-action table to perform the corresponding action.
For example, the software can follow a row of \actionShift{f5}{GB}{-50}{-50} to create a frame by shifting the green-blue channels of frame \codef{f5} 50 pixels up and left.
Alternatively, the software may visualize the generated actions' effects (\eg, the frames to create) and wait for the user's confirmation to enable an interactive PBE.

\smalltitle{Contributions}
This paper presents a novel meta DSL \lang that supports table-wise operations and common value computations.
The former includes common SQL operations such as filter, aggregation, join, and union.
The latter includes string manipulations and integral arithmetics.
To boost efficiency, the synthesis in \proj is driven by a bidirectional algorithm.
The forward direction follows the syntax-guided search to explore table-wise operations.
The backward direction decomposes the output table into sub-tables (sub-synthesis-problems), solves each separately, and merges the solutions later.
The interplay between two search directions synthesizes the value computations by solving constraints deduced from matching forward generated tables and backward propagated sub-tables.

We evaluated the effectiveness of \proj by comparing it with synthesizers implemented atop \prose, a state-of-the-art PBE framework, on 64 benchmarks from three application domains: file management, spreadsheet transformation, and XML transformation.
The domain-specifically optimized synthesizers powered by \prose solved 47 out of 64 benchmarks,
while \proj solved 53 out of 64.
The results indicate that the bidirectional synthesis algorithm of \proj is effective for practical scenarios. \proj achieves a performance comparable to that of the domain-specifically optimized synthesizers.

We also evaluated the usability of \proj via a human study with 12 participants without a program synthesis background.
They were asked to complete PBE tasks with both \proj and \prose.
The feedback shows that \proj is significantly more accessible than \prose in developing PBE tools.

In summary, this paper makes four major contributions.
\begin{enumerate}
\item We propose a customizable PBE framework named \proj that allows programmers to implement PBE tools for their software without expertise in program synthesis.
Many existing PBE applications~\cite{visual19, extract14, extract16, extract17, edit19, table11, file16} can be viewed in this data-action manner.

\item We adopt relational tables and SQL-like language \lang to model PBE tasks and design a bidirectional synthesis algorithm that can effectively guide the search over the large search space specified by \lang.

\item We implement a prototype of \proj and apply it to three data processing domains, namely, file management, spreadsheet transformation, and XML transformation.
\item We evaluate \proj in terms of usability and performance.
The evaluation results show that programmers find it effective to implement PBE tools with \proj,
and \proj's algorithm is capable of handling real-world PBE tasks.
\end{enumerate}

%% file: 2_overview.tex
\section{Overview}\label{sec:overview}
This section takes the GIF editing task in \Cref{fig:task-io} as an example to
illustrate the workflow of (1) how PBE developers can build applications on \proj, and (2) how \proj synthesizes target programs.
In \Cref{subsec:dev-interface}, we demonstrate how developers can easily leverage the interface to specify the figures and user actions as tables with several lines of code.
In \Cref{subsec:synth-proc}, we introduce the programs synthesized by \proj to automate PBE tasks
and briefly illustrate our bidirectional approach to synthesizing programs.

\subsection{Developer Interface}\label{subsec:dev-interface}
\proj relieves developers from designing DSLs by using a predefined meta DSL to process entities and actions in various domains.
To apply \proj to PBE tasks in a specific domain (\eg, GIF editing in \Cref{fig:task-io}), developers need only to model
data entities and user actions in the domain as relational tables.
Relational tables in \proj are ``flat'' in that only primitive types (\eg, integers and strings) are allowed. %
The design decision follows Daniel Jackson's Alloy~\cite{alloy-book}, which points out that
flat tables are sufficiently general for software abstractions and tractable for search-based software automation.

\smalltitle{Data Entities}
To use \proj, developers should provide a data entity \emph{model} that
(1) identifies what data (and data fields) are related to the repetitive tasks aiming to be automated, and
(2) represents the identified data (and data fields) as relational tables.
Data entities can be identified from the subjects in the repetitive tasks, \ie, what the users usually need to operate frequently.
For example, when editing GIFs, the frame is where users apply changes and thus a natural choice for the data entity.
Developers also need to decide which data fields of the identified data entities are needed in the repetitive tasks.
For example, a frame's order in the frame sequence is needed because some PBE tasks (\eg, our motivating task) need to perform user actions dependent on the order.

Modeling the data entities for PBE tasks in tables is like modeling a relational database.
A row and a column in the table represent one data entity and one identified data field, respectively.
For instance, the four rows in the Input Example table in \Cref{fig:task-io} represent four frame entities,
and the columns (\eg, order in the frame sequence) represent various fields of the frames.
\proj allows each data entity to be indexed by a unique \emph{identifier} (\eg, the \codef{id} column in \Cref{fig:task-io}).

Implementation of a table schema can be accomplished through \proj's ORM-style~\cite{orm} APIs.
\Cref{lst:data-entities} gives an example of implementing the schema for the Input Example table in \Cref{fig:task-io}.
In a nutshell, a table can be implemented by a class annotated with \codef{[Entity]} (annotations in C\#).
Table columns are realized by methods annotated with \codef{[Field]}.
Specially, methods annotated with \codef{[IdField]} return \codef{[Entity]} objects, where each \codef{[Entity]} object would be automatically converted to a unique value in identifier columns (\eg, the \codef{id} column in \Cref{fig:task-io}).

\input{codes/data_example}

\input{codes/action_example}

\smalltitle{User Actions}
While data entities describe a system's state, user actions over entities can describe state changes.
Like data entities, user actions can be modeled as relational tables.
Without loss of generality,
we assume that (1) developers hold a set of predefined user actions that can be taken to change the system's state, such as renaming/moving files or making various effects as in \Cref{fig:task-io};
(2) each action can be modeled by a command \codef{action(arg1, arg2, ...)}.%
\footnote{The number of arguments depends on the action, \eg, the directory creating action \codef{mkdir(path)} has one argument while the file moving action \codef{mv(old\_path, new\_path)} has two. }
\proj generates the table representation of user actions by taking \codef{action}, \codef{arg1}, $\ldots$ as columns and each action instance as a row.
Therefore, developers only need to define user actions in the form of commands (action name and arguments).

User actions are usually implemented as functions/methods over entities.
Thus, \proj allows specifying actions with annotations on methods, as the C\# example for the shift action in \Cref{lst:user-actions}.

\smalltitle{PBE Tasks}
In the scenario of automating repetitive user actions over data entities,
the process of producing the desired user actions from the inputted data entities usually follows a task-specific logic, \ie, the target program in PBE systems.

Such programs can usually be generalized from a small set of example inputs (data entities) and example outputs (user actions).
Each PBE task is a problem of generating programs consistent with the given input-output examples.
In \proj, the workflow of solving a PBE task is as follows:
\begin{enumerate}
    \item Users initiate a PBE task (\eg, by pressing a ``start PBE'' button) and select example data entities (\eg, the first several frames in \Cref{fig:task-io}).
    \item The software application passes the corresponding \codef{[Entity]} objects to \proj, and \proj converts them into tables.
    \item Users perform example actions (\eg, setting graphic effects in \Cref{fig:task-io}).
    \item Each user action triggers an invocation of an \codef{[Action]} method. \proj records the arguments and converts them into tables.
    \item \proj synthesizes programs consistent with the example inputs and outputs.
\end{enumerate}

In summary, using \proj mainly requires the specification of data entities and user actions without knowledge of the domain-specific language and program synthesis algorithm.
Such a PBE workflow can be integrated into software applications via lightweight engineering efforts.
Further details of the \proj developer interface are available in our tutorial~\cite{beehomepage}, including three PBE tools as instances.

\subsection{\proj Programs and Synthesis Procedure}\label{subsec:synth-proc}

\smalltitle{\proj Programs}
In \proj, the process of generating user actions from data entities corresponds to a program performing a sequence of ``intermediate actions'', each creating a new intermediate table.
The final intermediate table contains the user actions.
For example, the logic for the task in \Cref{fig:task-io} can be decomposed mechanically into two processes:
(1) Extract odd and even frames into two separate tables that compute the biases to shift differently;
(2) Generate a user action with the computed biases for each frame.
Therefore, \proj programs are designed to consist of two phases accordingly.

\begin{enumerate}
\item The \emph{transformation} phase applies a sequence of table-to-table transformations (\eg, \opFilter, \opJoin) to generate intermediate tables that would participate in the next phase.
\item The \emph{mapping} phase generates a table of user actions by projecting the intermediate tables onto the target columns (computed from expressions) using \opYield statements.
\end{enumerate}

\begin{figure}
    \begin{center}
        \includegraphics[scale=0.4, page=1]{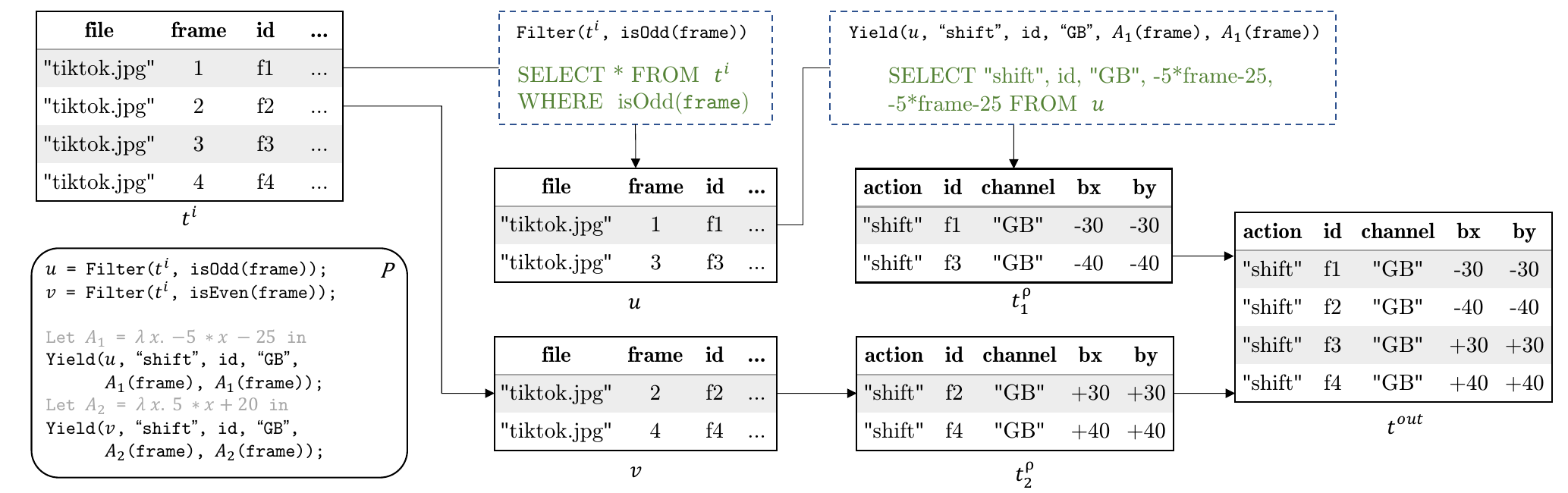}
    \end{center}
        \caption{The program $\symprog$ for the task in \Cref{fig:task-io} and example execution procedure with informal semantics in SQL.
        We use ``Let'' expressions here to introduce $A_1$ and $A_2$ to simplify the presentation (the formal grammar of \proj does not include ``Let'').
    }
    \label{fig:execution}
\end{figure}

The transformation phase can be written as a series of SQL-like operators (e.g., \opFilter resembles a \texttt{SELECT} with a \texttt{WHERE} clause) that take in a table together with a predicate and output an intermediate table.
\Cref{fig:execution} shows the synthesized program $\symprog$ that generalizes the given input/output examples.
The first two statements are \opFilter statements.
The first (resp. second) statement takes in the table $t^i$ (example frames) along with a predicate judging if the frame is odd (resp., even) and outputs a table $u$ (resp., $v$) containing odd-frames (resp., even-frames).

The mapping phase can be written as a sequence of \opYield statements.
Each \opYield statement projects a table onto some given ``columns'' like the \texttt{SELECT} clause in SQL.
In addition, we allow arithmetic or string expressions in the projection to support computations such as the bias shifts.
Take the first \opYield statement in \Cref{fig:execution} for example.
It takes in table $u$ (odd-frames) and outputs table $\tpja$ by executing \opYield in the following ways:
\begin{enumerate}
\item Constant expressions (\eg, ``shift'' and ``GB'' in \opYield) result in columns containing the constant values.
\item Expressions containing a column name (\eg, \codef{id} in \opYield) result in the projected column.
\item Expressions with computations (\eg, $A_1(\codef{frame})$ in \opYield) result in columns that contain the computed values for each row in the input table.
Take $A_1(\codef{frame}) = -5 * \codef{frame} - 25$ for example, the resulting column contains $-30$ computed from the first row in $u$ ($\codef{frame}=1$) and $-40$ computed from the second row ($\codef{frame}=3$).
\end{enumerate}
Moreover, tables generated from \opYield statements (\eg, $\tpja$ and $\tpjb$) would be unioned into a single table (\eg, $\tout$) as final output (the desired user actions).

\smalltitle{\proj Synthesis Algorithm}
To obtain a \lang program like the one in \Cref{fig:execution}, we need to synthesize both the transformation part (\eg, the two \opFilter statements) and the mapping part (\eg, the two \opYield statements).
A simple algorithm to synthesize example-consistent \lang programs is to exhaustively enumerate all syntactically valid programs (\Cref{subsec:language}) and filter out those that are inconsistent with the given examples.

However, such a simple syntax-guided search strategy is unscalable.
There could be hundreds of thousands of programs (and resulting tables) in the transformation phase to enumerate, even if only considering two to three transformation statements.
The expressions with constant parameters (\eg, $-5$ and $-25$ in $A_1$) and the ``union'' semantics of $\opYield$ statements make efficient searching even more challenging.
This paper presents a bidirectional search algorithm in which the backward direction generates partial programs with placeholders whose exact values (\eg, parameters in computation expressions) are determined by a matching phase.

\begin{figure*}
    \begin{center}
        \includegraphics[scale=0.35, page=1]{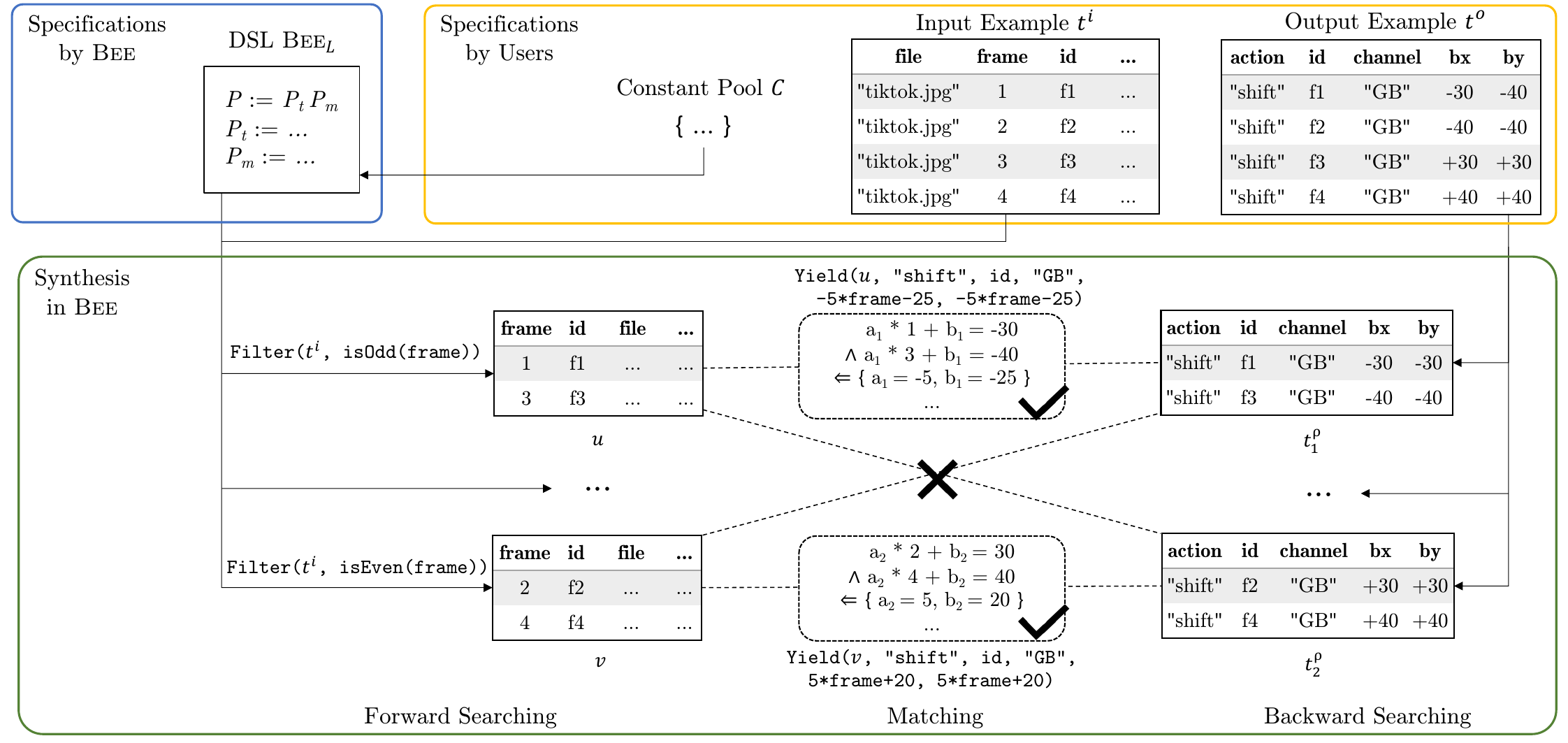}
    \end{center}
    \caption{The bidirectional synthesis procedure for the program in \Cref{fig:execution}.
    The specifications to the synthesis task include the part provided by \proj (DSL) and the parts provided by users through the software application (input/output example tables and a constant pool).
    The synthesis procedure in \proj includes forward searching, backward searching, and matching between the two directions.
    The constant pool $\constpool$ in this example is empty.}
    \label{fig:synthesis}
\end{figure*}

Specifically, statements in the transformation phase are synthesized in the forward-searching by enumerating possible table-to-table transformation statements (\eg, \opFilter, \opJoin) according to the grammar.
\Cref{fig:synthesis} shows the synthesis process for the program in \Cref{fig:execution},
where the two \opFilter statements generating $u$ and $v$ are enumerated in the forward-searching.

\opYield statements in the mapping phase are synthesized in the backward-searching.
The backward searching first enumerates all possible sub-tables of the example output table, each corresponding to the output of a \opYield statement.
For example, tables $\tpja$ and $\tpjb$ are enumerated in \Cref{fig:synthesis}.
Considering that the number of sub-tables is exponential to the number of rows in the output table,
we need to rank the sub-tables to avoid explosion when the number of rows is large (detailed in~\Cref{subsec:optimizations}).

The expressions in the \opYield statements are synthesized by matching
forward searched intermediate tables against backward searched tables and solving constraints.
Specifically, we first range over a forward table, a backward table, and their row-to-row mapping.
Then, for each column in the backward table,
we enumerate a parameterizable expression from a predefined set that includes common arithmetics and string manipulations.
Finally, we check (\eg, with an SMT solver) if there exist parameters for the expression that exactly maps some columns in the forward table to the backward table.

For example, consider that we are matching table $u$ and table $\tpja$ in \Cref{fig:synthesis} and we know the row-to-row mapping by comparing the \codef{id} columns.
To synthesize column \codef{bx} that needs arithmetic computation,
we first enumerate the parameterizable expression $\lambda x. a * x + b$,
then check that the \codef{frame} column of $u$ can be mapped to column \codef{bx} with ${a = -5, b = -25}$.

To summarize, with bidirectional searching and constraint solving, we avoid enumerating parameters in expressions, which is infeasible concerning synthesis efficiency.
We present the complete algorithm in \Cref{sec:algorithm}.

%% file: codes/data_example.tex
\begin{lstlisting}[language={[Sharp]C}, label={lst:data-entities},caption={A C\# code snippet for specifying the table schema of Input Example in \Cref{fig:task-io}.}]
[Entity]
public class GraphicEntity {
    private File _file;
    private Frame _frame;
    GraphicEntity(File file, Frame frame) {
        _file = file;
        _frame = frame;
    }

    [Field]
    public string File() {
        return _file.Filename;
    }

    [Field]
    public int Frame() {
        return _frame.order;
    }

    [IdField]
    public GraphicEntity Id() {
        return this;
    }

    ...
}\end{lstlisting}

%% file: codes/action_example.tex
\begin{lstlisting}[language={[Sharp]C}, label={lst:user-actions},caption={A C\# code snippet for specifying the shift action, \ie, the table schema of Output Example in \Cref{fig:task-io}.}]
[Entity]
public class GraphicEntity {
    ...

    [Action]
    public static void Shift(GraphicEntity ge, string rgb, int biasX, int biasY) {
        foreach (var channel in ge._frame.GetChannels(rgb))
            channel.translate(biasX, biasY);
    }
}\end{lstlisting}

%% file: 3_specification.tex
\section{The \proj Specification}\label{sec:formulation}
With the developer-provided modeling of data entities and user actions,
as well as the workflow introduced in \Cref{subsec:dev-interface},
solving a PBE task in \proj is like invoking a function $\synthfunc (\tuple{\texi, \texo}, \tpend, \constpool)$,
where $\synthfunc$ is the program synthesizer in \proj,
$\texi$ denotes the tables converted from example data entities,
$\texo$ denotes the table converted from example user actions,
$\tpend$ denotes the table converted from data entities pending processing,
and $\constpool$ is a constant pool.

In this section, we first formulate the above synthesis problem in \Cref{subsec:interface-spec},
then introduce our meta DSL \lang that defines the synthesizable programs in \proj (\Cref{subsec:language} to~\ref{subsec:lang-program}).

\subsection{Synthesizer Interface Specification}\label{subsec:interface-spec}

\smalltitle{Tables}
Two-dimensional table representation of domain-specific data/action is the foundation of \proj.
The inputs $\texi, \texo, \tpend$ to \synthfunc are all based on tables.
Formally, an $n$-ary table $t$ is an \emph{unordered} set of typed tuples.
Each table $t$ is associated with a name $\symname$ and a table \emph{schema}
\[\symschema = \langle col_1:\tau_1, \ldots, col_n:\tau_n \rangle,\]
where $col_i$ and $\tau_i$ denote the $i$-th column's name and type, respectively.
Each row $\row \in t$ is a tuple $\langle v_1, \ldots, v_n \rangle$, where each $v_i$ is of type $\tau_i \in \{\type{Int}, \type{String}, \typeid\} $.
\typeid, like the primary key in SQL, uniquely distinguishes entities in a table.
The following operations are defined over tables:

\begin{itemize}
    \item (\emph{fetch}) $t[\row, \col]$ or $\row[\col]$ ($t$ is omitted when the context is clear) denotes the value of the cell at $\row$ and $\col$ in table $t$.
    \item (\emph{project}) $t[\col_1,\ldots,\col_n]$ denotes a new table obtained by removing columns other than $\col_1,\ldots,\col_n$ in $t$ and deduplicating repetitive rows.
    \item (\emph{append}) $t \append \col_{new} $ denotes appending $\col_{new}$ (a list of values) to $t$.
    Suppose $t$ is an $n$-ary table with $m$ rows $\row_1, \ldots, \row_m$, and $\col_{new}$ is a list of values $v_1, \ldots, v_m$.
    $t \append \col_{new} $ yields a new $(n\! +\! 1)$-ary table with $m$ rows, where the $i$-th row is an $(n\! +\! 1)$-ary tuple appending $v_i$ to $\row_i$.
    \item (\emph{union}) Suppose $t_1$ and $t_2$ have the same table schema $\symschema$, $t_1 \cup t_2$ denotes a new table with its schema being $\symschema$ and its rows being the union of rows in $t_1$ and $t_2$.
\end{itemize}

\smalltitle{Inputs to \synthfunc}
The inputs to $\synthfunc (\tuple{\texi, \texo}, \tpend, \constpool)$ comprise the followings:

$\texi$ (\emph{input examples}) is a list of input examples (tables) that model the application domain's data entities.
In $\texi = \tuple{t_1, \ldots, t_n}$,
each table $t_x$ is associated with a table name $\symname_x$ and a table schema $\symschema_x$.
Data entities can be modeled by table schemas, \eg, graphic frames can be modeled with columns \codef{file}, \codef{frame}, \codef{id}, \etc., as in \Cref{fig:task-io}.
Entity relations can also be modeled by table schemas, \eg, the parental relation in XML documents can be modeled by
$\symschema = \tuple{ \textrm{parent}:\typeid,\ \textrm{child}:\typeid }$,
where each pair denotes a direct inclusion relation between two XML elements.

\smallskip
$\texo$ (\emph{output examples}) models the application domain's example actions performed on $\texi$.
$\texo$ has a fixed schema
\[\symschemao = \tuple{\textrm{action}:\typestr, \ \mathit{arg}_1:\tau_1,\ \ldots}.\]
Specifically, the first column denotes the kind of action to be performed.
The remaining columns denote the arguments of the action.
For example, if the action is ``rename'', the first argument should be the file path at present, and the second argument should be the new file path.
For concise presentation, this paper assumes the $\texo$ in each invocation of $\synthfunc$ has only one kind of action.
In case multiple kinds of user actions were performed (\eg, file removal and renaming), \synthfunc can be invoked independently for each kind of action.

\smallskip
$\tpend$ (\emph{pending data}) is a list of tables that are the additional (non-example) inputs that need to be processed by the synthesized program.
In $\tpend = \tuple{t_1, \ldots, t_n}$, each $t_x$ has the name $\symname_x$ and schema $\symschema_x$.
Note that $\tpend$ should have the same structure as that of the input example $\texi = \tuple{t'_1, \ldots, t'_n}$,
i.e., for any $x$, $\symname_x = \symname'_x$ (name of $t'_x$) and $\symschema_x = \symschema'_x$ (schema of $t'_x$).
This assures that the program synthesized from $\tuple{\texi, \texo}$ can be applied to $\tpend$.

\smallskip
$\constpool$ (\emph{constant pool}) is a set of constant values (can be empty) as ingredients of predicates synthesized by \synthfunc.
Our motivating task does not require such constants.
Take another task as an example.
If we want to delete \codef{pdf} files, ``pdf'' should be included in $\constpool$.
The availability of a constant pool facilitates the synthesis procedure, typically reducing wrong search branches when generating ``where'' clauses.
As observed from online forums and suggested by previous SQL synthesis work~\cite{sql17}, the constants can be commonly identified from the textural task description.
Hence, the software application may ask the end-users to provide such ``keywords'' in prompt or automatically extract from essential fields in the application domain (\eg, file extension in file management).

\smalltitle{Synthesis Problem}
The output of $\synthfunc(\tuple{\texi, \texo}, \tpend, \constpool)$ is a program $\symprog$ written in a language called \lang such that
\begin{equation*}
    \eval{\symprog(\texi)} = \texo,
\end{equation*}
i.e., $\symprog$ is the generalization of the input-output example $\tuple{\texi, \texo}$.
Applying $\symprog$ to the pending data $\tpend$ yields the target user actions $\ttarget = \eval{\symprog(\tpend)}$.

\subsection{\lang Syntax and Semantics}\label{subsec:language}
\lang is the meta DSL defining the exact scope of \proj, \ie, the set of programs that can be synthesized to solve PBE tasks.

\smalltitle{Syntax}
The syntax of \lang is defined in \Cref{fig:syntax}.
\lang is designed based on adding value computation to SQL and aims to reach a balance between practical expressiveness and synthesis tractability.
A \lang program $\symprog$ consists of two parts:
a transformation program $\sympt$ for creating useful intermediate tables followed by a mapping program $\sympm$ for projecting intermediate tables to actions.
Constants in predicate $\pred$ are from the constant pool $\constpool$, while other constants can be inferred automatically.

\begin{figure}
    \centering
    \begin{tabular}[t]{rrcl}
        Program  & $\symprog$ & $\coloneqq$ & $\sympt \ \sympm$ \\
        \midrule
        Transform Program & $\sympt$ & $\coloneqq$ & $\symst$; $\sympt$ \altab $\epsilon$ \\
        Transform Statement & $\symst$ & $\coloneqq$ & $t$ = \opFilter ($t$, $\pred$)\\
        & & \altab & $t$ = \opJoin ($t_1$, $t_2$, $\col_{1}$, $\col_{2}$)\\
        & & \altab & $t$ = \opGroup ($t$, $col_{index}$, ($\aggr$, $col$)$\ldots$)\\
        & & \altab & $t$ = \opOrder ($t$, $col$, $c_{start}$, $c_{inv}$)\\
        & & \altab & $t$ = \opOrder ($t$, $col$, $c_{start}$, $c_{inv}$, $col_{index}$)\\
        Aggregation & $\aggr$ & $\coloneqq$ & max \altab min \altab sum \altab avg \altab cnt \\
        Predicate & $\pred$ & $\coloneqq$ &  ($\pred_1 \land \pred_2$) \altab ($\pred_1 \lor \pred_2$) \altab $\neg \pred$ \\
        & & \altab &  $ps$ ($col_1, col_2$)  \altab  $ps$ ($col, c$) \altab  $ps$ ($col$) \\
        \midrule
        Mapping Program & $\sympm$ & $\coloneqq$ & $\symsm$; $\sympm$ \altab $\epsilon$ \\
        Mapping Statement & $\symsm$  & $\coloneqq$ & \opYield $(t, \projexpr\ldots)$ \\
        Projection & $\projexpr$   & $\coloneqq$ & $col$  \altab  $c$ \altab \opMutate ($f$, $col\ldots$) \\
    \end{tabular}
    \caption{The context-free grammar of \lang. $\epsilon$ denotes an empty literal,
        $f$ denotes a feature (introduced later),
        $t$ denotes a table variable,
        $col$ denotes a column name,
        $c$ denotes a constant value,
        $ps$ denotes a predicate symbol.}
    \label{fig:syntax}
\end{figure}

\smalltitle{Semantics of Transformation Program}
$\sympt$ is a sequence of transformation statements.
Each transformation statement $\symst$ generates a new table variable by applying a transformation on existing tables.
The following transformations are defined in \lang:

\construct{\opFilter} resembles the ``where'' clause \ul{\texttt{SELECT * FROM $t$ WHERE $\pred$}} in SQL for applying predicates on rows.
Each predicate symbol $ps$ in $\pred$ receives one to two columns/constants and produces a boolean value, \eg, \texttt{IntEquals}$(\col_1, \col_2)$, \texttt{IsSubstring}($\col, c$), \texttt{IsOdd}($\col$).
Constants $c$ are picked from the end-user-provided constant pool $C$ (recall that $\constpool$ is input to \synthfunc).

\construct{\opJoin} resembles the ``join'' clause in SQL for connecting two tables: \ul{\texttt{SELECT * FROM $t_1$ JOIN $t_2$ ON $t_1.\col_1=t_2.\col_2$}}.
Unlike SQL, \lang only allows joining two columns of \typeid type.
We made such a tradeoff because:
(1) Eliminating joining on columns of arbitrary type (\eg, strings) significantly reduces the search space.
(2) We observe that most join operations needed in practical PBE tasks belong to two categories, one is joining on \typeid columns, and the other is covered by the \opGroup operator introduced below.

\construct{\opGroup} aims to resemble the ``group'' clause in SQL for aggregating values, \eg, \ul{\texttt{SELECT $\aggr_1 \ \col_1, \dots$ FROM $t$ GROUP BY $\col_{index}$}}.
In practice, the semantics of \opGroup is adjusted because we notice a common type of PBE task that needs to join the aggregated values and the values in the original table.
A typical example is to extract the filename (original data) of the largest file under each folder (aggregation).
To save a join step after aggregation, instead of only producing the aggregated table, \opGroup produces a table by joining the original table and aggregated table on the index column.
That is, it resembles \ul{\texttt{SELECT * FROM $t$ JOIN $g = $ (SELECT $\aggr_1 \ \col_1, \dots$ FROM $t$ GROUP BY $\col_{index}$}) ON $t.\col_{index} = g.\col_{index}$} in SQL.
Each newly generated column by aggregation is assigned a new name distinct from existing columns.

\construct{\opOrder} resembles the ``order'' clause
\ul{\texttt{SELECT * FROM $t$ ORDER BY $\col$}} in SQL for sorting rows.
However, tables in \lang are unordered sets and cannot be sorted.n
Alternatively, \lang appends a new column to indicate each row's rank sorted by \opOrder.

\begin{figure}
    \begin{center}
        \begin{subfigure}[b]{0.2\textwidth}
            \centering
            \includegraphics[page=1,scale=0.25]{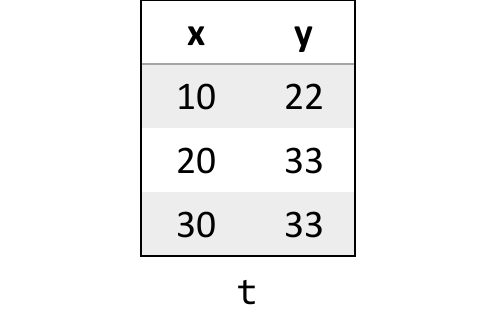}
            \caption{Origin}
            \label{fig:origin}
        \end{subfigure}
        \hfill
        \begin{subfigure}[b]{0.2\textwidth}
            \centering
            \includegraphics[page=3,scale=0.25]{fig/fig_language}
            \caption{Order}
            \label{fig:order}
        \end{subfigure}
        \hfill
        \begin{subfigure}[b]{0.2\textwidth}
            \centering
            \includegraphics[page=5,scale=0.25]{fig/fig_language}
            \caption{Order with index}
            \label{fig:order_index}
        \end{subfigure}
        \hfill
        \begin{subfigure}[b]{0.2\textwidth}
            \centering
            \includegraphics[page=2,scale=0.25]{fig/fig_language}
            \caption{Mutate}
            \label{fig:mutate}
        \end{subfigure}
    \end{center}
    \caption{Examples of operators \opOrder and \opMutate.}
    \label{fig:mutate-order}
\end{figure}

An example of using \opOrder is shown in \Cref{fig:order}, where column \texttt{ord} is ordered on column \texttt{x}.
If an optional column argument $\col_{index}$ is provided, the values in $\col$ only compare with rows of the same value at $\col_{index}$.
An example of using such an index column is shown in \Cref{fig:order_index}.
Different from \Cref{fig:order}, the new value at row 2 is 0 because only row 3 shares the same value with row 2 at column \texttt{y}.
The new value at row 3 becomes 1 for the same reason.

\smalltitle{Semantics of Mapping Program}
The mapping program $\sympm$ consists of a sequence of \opYield statements and finally produces the table $\tout$ of actions.
Suppose the $k$-th \opYield statement returns table $t_k$, $\tout$ is the union of all \opYield results, i.e.,
\[\tout = \bigcup t_k.\]

\construct{Yield} takes in a table $t$ and multiple projection expressions $\projexpr$.
Each $\projexpr_i$ is evaluated as a column $\col_i$ before executing \opYield, and the evaluation of \opYield projects $t$ on columns $\col_i$, i.e.,
\[\eval{\text{\opYield}(t, \pj_1, \ldots, \pj_n)} = t[\col_1, \ldots, \col_n].\]

Each projection expression $\projexpr$ computes a column to be projected in \opYield.
Three types of projections are allowed:

\smallskip
\begin{itemize}
    \item Column name $\col$ is evaluated to be the column $\col$ in $t$.
    \item Constant $c$ is evaluated to be a column filled with constant $c$.
    Typically, the first $\pj$ expression in a \opYield statement must be a constant string denoting the action name, \eg, ``shift'' in \Cref{fig:execution}.
    \item \opMutate expression is evaluated to be a column, as explained below.
\end{itemize}

\smallskip
\opMutate is for the case where the PBE task may involve domain-specific computations, e.g., 
arithmetic computations are required for the task in \Cref{fig:task-io}.
However, these computations are beyond the expressiveness of the commonly used SQL operators (\eg, where, join).
To strike a balance between expressiveness (\eg, allowing such computations everywhere, even in $\sympt$) and synthesis tractability,
we limit such computation in \opMutate at the mapping phase.

Specifically, \opMutate takes in a \emph{feature} $f$, \ie, a function of type $\tau_1 \times \ldots \times \tau_n \to \tau_o$,
a list of selected columns $\col_1:\tau_1, \ldots, \col_n:\tau_n$,
and an implicit table $t$ in the \opYield statement.
The mutation is conducted for each $\row$ of $t$,
yielding a new $\tau_o$-typed column $\col_{new}$, denoted by $f(\col_1, \ldots, \col_n)$, where the value in each row $\row$ is computed by
\[\row[\col_{new}] = f(\row[\col_1], \ldots, \row[\col_n]).\]
For example, in \Cref{fig:mutate}, the feature $f$ is $\texttt{sum}(100)$ that sums two values plus a bias of 100,
and the column \texttt{sum} is obtained by applying $f$ on columns \texttt{x} and \texttt{y}.

To make \lang applicable to a broad spectrum of PBE tasks, we introduce \emph{higher-order features},
\ie, functions that take parameters as input and output a feature.
\lang incorporates the following higher-order features.

\newcommand{\outputtype}[1]{{\color{gray}\footnotesize #1}}
\begin{equation}
    \footnotesize
    \begin{aligned}
        \texttt{substring}(regex) &= \lambda x.\ \text{extract}(x, regex): \outputtype{\typestr {\to} \typestr} \\
        \texttt{concat}(M)        &= \lambda x_1 \dots \lambda x_n .\ M(x_1, \dots, x_n): \outputtype{\typestr^n  {\to} \typestr} \\
        \texttt{sum}(b)           &= \lambda x. \lambda y.\ x+y+b: \outputtype{\typeint \times \typeint {\to} \typeint} \\
        {\texttt{linear}}(a,b)    &= \lambda x.\ ax+b: \outputtype{\typeint {\to} \typeint} \\
        {\texttt{div}}(b,d)       &= \lambda x.\ (x+b)\ \text{div} \ d: \outputtype{\typeint {\to} \typeint} \\
        {\texttt{mod}}(b_1,b_2,d) &= \lambda x.\ (x+b_1) \ \text{mod} \ d + b_2: \outputtype{\typeint {\to} \typeint} \\
    \end{aligned}
    \label{eq:features}
\end{equation}

These higher-order features are useful in abstracting common PBE tasks and are well-supported by modern constraint solvers~\cite{z3} or program synthesizers~\cite{flashfill11}.
For example, \texttt{div} and \texttt{mod} can be used to plot a sequence of $3 \times 2$ blocks to a sequence of $6 \times 1$ blocks.
~\footnote{The \texttt{mod} feature only supports a fixed range of common dividends (2 to 10) since an unlimited range goes beyond a constraint solver's capability.}
\texttt{concat} can simulate a lot of string-like transformations.
For example, the movement of files can often be viewed as path string manipulation.

The \texttt{substring} and \texttt{concat} features are for string manipulation.
An \text{extract}($x, regex$) expression extracts a substring of $x$ specified by a regular expression $regex$,
where the scope of $regex$ is borrowed from FlashFill~\cite{flashfill11}.
A string manipulation program $M$ is characterized by $m$ regular expressions,
and $M(x_1, \ldots, x_n)$ concatenates (denoted by $\oplus$) $m$ substrings:

\begin{equation}
    \small
    \begin{aligned}
        M(x_1, \ldots, x_n) \ &= \ \text{extract}(c_1, regex_1) \oplus \ldots \oplus \text{extract}(c_m, regex_m),\\
        \text{where} \        & c_i \in \{ x_1, \ldots, x_n \}.
    \end{aligned}
    \label{equ:string-manipulate-program}
\end{equation}

In our workflow, these higher-order features are designed to be built in \proj instead of provided by PBE developers.
First, solving the feature parameters requires sound background knowledge of SMT and PBE. The requirement violates our goal to make \proj easy to use.
Second, \proj's library of higher-order features can be extended by PBE experts, while the users only need to select those they need.
In principle, a high-order feature can be added as long as there exists an efficient backbone (\ie, constraint solver or program synthesizer).

\subsection{Execution and Validity of \lang Programs}\label{subsec:lang-program}

$\synthfunc(\tuple{\texi, \texo}, \tpend, \constpool)$ produces a program $\symprog$ such that $\eval{\symprog(\texi)} = \texo$.
\synthfunc only considers \emph{valid} programs.
A \lang program is considered valid if no type violation occurs, and all used variables are defined before use during execution.

Specifically, the validity check is done statement by statement during execution.
The execution of $\sympt$ uses a set $\symstate$ to maintain the tables that have been defined.
Suppose that $\symprog(\langle t_1, \ldots, t_n \rangle)$ is executed, $\symstate$ is initialized as $\symstate = \{ t_1, \ldots, t_n \}$.
Each execution of a transformation statement $t = op(\ldots)$ adds the generated table $t$ to $\symstate$.
A transformation statement $t = op(\ldots)$ is valid if the followings hold.
\begin{enumerate}
    \item The table variables and column variables used in $\symst$ are defined in $\symstate$.
    \item The name of the generated table $t$ does not duplicate with the table names in $\symstate$.
    \item The operations in $op(\ldots)$ are type-checked, \eg, \opJoin requires the two columns to have the same type.
\end{enumerate}

A \opYield statement $\text{\opYield}(t, \pj_1, \ldots, \pj_n)$ is valid if the followings hold.
\begin{enumerate}
    \item $t$ and the column variables used in $\pj_1, \ldots \pj_n$ are defined in $\symstate$.
    \item $\pj_1$ is a constant string that corresponds to a defined action $a$.
    \item The operations in $\pj_1, \ldots \pj_n$ are type-checked.
    \item The types of $\pj_1, \ldots \pj_n$ are consistent with the action arguments of $a$.
\end{enumerate}

%% file: 4_algorithm.tex
\section{The \proj Synthesizer}\label{sec:algorithm}

This section illustrates our approach to synthesizing a \lang program $\symprog$ satisfying the given input-output example $\texi$ and $\texo$, \ie, the synthesis problem defined in \Cref{subsec:language}.
Before diving into the algorithm, we first introduce the following challenges under our problem setting if we were to
adopt the syntax-guided enumeration-based synthesis algorithm commonly used in SQL query synthesis~\cite{sql09, sql13, sql17}.

\subsection{Motivation}\label{subsec:algo-motivation}
$\symprog$ consists of
$\sympt$, a sequence of table operations to create intermediate tables,
and $\sympm$, a sequence of projections.
Therefore, the straightforward baseline algorithm could be:

\begin{enumerate}
  \item Synthesizing $\sympt$ by maintaining a work list of tables $T$ (initially, $T = \texi $).
  Each search iteration enumerates all possible \opFilter/\opJoin/\opGroup/\opOrder operations on tables in $T$ (\opJoin operates on $(t, t') \in T\times T$) and adds the result tables back to $T$.
  \item To synthesize $\sympm$, we can synthesize each \opYield statement by deciding whether each $t \in T$ can be projected onto a subset of $\texo$, \ie, whether $\exists \pj_1, \ldots, \pj_n.$ such that $\eval{\text{\opYield}(t, \pj_1 \ldots \pj_n)} \subseteq \texo$.
  Then, $\sympm$ can be obtained by deciding whether there exists a set of \opYield statements such that the union of their projected tables is $\texo$.
\end{enumerate}

This baseline algorithm is inefficient because
we need to synthesize features' parameters in \Cref{eq:features}, \eg, regular expression $regex$ in \texttt{substring}.
Brute-force enumeration of parameters (integers and regular expressions) yields a huge search space.
Alternatively, a practical approach is to deduce the parameters (\eg, via a solver) given the values this feature is supposed to output.
However, this requires enumerating all possible subsets of $\texo$ because the subset's values would decide the deduction results of the parameters.
For example, when synthesizing the feature in \Cref{fig:execution},
among the 15 non-empty subsets, only \{\textrm{-30, -40}\} and \{\textrm{+30, 430}\} can deduce the correct linear arithmetic expression $A_1$ and $A_2$, respectively.
In the worst case, we need to explore $O(2^n)$ subsets of $\texo$, where $n$ is the number of rows in $\texo$ ($n$ can be up to 30 in our evaluation benchmarks).
Note that such worst-case frequently happens because the table $t$ we try to find projection ranges over all the intermediate tables in $T$.
Many of them do not have projections but still require trying with all the subsets of $\texo$.

In response to this challenge, we observe that we can alternate the search order to \emph{hypothesize} a sub-table of $\texo$ (a subset of rows, denoted by $h \subseteq \texo$) in advance.
With effective prioritization of sub-table search (detailed in \Cref{subsec:optimizations}),
the synthesis of a large $\texo$ is divided into solving smaller sub-problems,
which greatly scales the problem size we can solve.
As reflected in the evaluation, this approach enabled \proj to synthesize programs for the tasks with 20 user actions,
while the baseline algorithm timed out for all tasks with more than ten user actions.

\subsection{Bidirectional Search}
The above observation derives a bidirectional search algorithm (\Cref{alg:top}) in our synthesizer:
\begin{enumerate}
  \item In the forward searching, we enumerate statements in $\sympt$ by syntax, and each one produces an intermediate table.
  \item In the backward searching, we prioritize sub-tables in $\texo$ and try to synthesize a \opYield statement for each of them.
  \item The higher-order feature parameters are determined during the table matching from two directions.
\end{enumerate}

\let\oldnl\nl%
\newcommand{\nonl}{\renewcommand{\nl}{\let\nl\oldnl}}

\begin{algorithm}[t] \small

\textbf{Function} Synthesize($\texi,\texo$) \\
\Begin{
$\ptset \gets \{ \tuple{\epsilon, t} \ | \ t \in \texi \} $\;
$\pmset \gets \emptyset$\;
\For{$d \gets 1 \ \textbf{to} \ \infty$}{
  $\ptset \gets \ptset \cup \Call{Expand}{\ptset, d} $\;
  $g \gets \Call{HypothesisGenerator}{\texi, \texo}$\;
  \While{$\Call{HasNext}{g}$}{
    $h \gets \Call{Next}{g}$\;
    $\symsm \gets \Call{Match}{h, \ptset}$\;
    \If{$\symsm \neq \bot$}{
      $\pmset \gets \pmset \cup \{ \tuple{\symsm, h} \} $\;
      $g \gets \Call{UpdateRank}{g, h}$\;
      $\sympm \gets \Call{AssembleMapping}{\pmset, \texo} $\;
      \If{$\sympm \neq \bot$}{
        \Return $\Call{AssembleProgram}{\ptset, \sympm}$;
      }
    }
  }
}
}

\caption{Top-level synthesis algorithm.}\label{alg:top}
\end{algorithm}

\smalltitle{Forward Searching}
The forward searching adopts the syntax-guided enumeration like the baseline algorithm in \Cref{subsec:algo-motivation}.
Specifically, we maintain a set $\ptset = \{ \tuple{\symst, t} \}$ where each element contains a transform statement $\symst$ and a result table $t$.
$\ptset$ is initialized with empty statements and tables from $\texi$ (line~3).

The forward searching enumerates statements in order of \emph{depth}.
We assign depth to each statement $\symst$ (and its corresponding result table $t$) as follows.
Empty statement $\epsilon$ and tables from $\texi$ have depth $0$.
The depth of a non-empty statement $\symst$ with operator $op$ and operand tables $\overline{t}$ is $max_{t \in \overline{t}}\text{depth}(t) + \text{depth}(op)$.
For operator \opFilter, \text{depth}{$(op)$} is the number of predicate symbols in the predicate, and for other operators, $\text{depth}$ is 1.

Specifically, the depth budget $d$ is increased in each iteration (line~5),
and the $\Call{Expand}{\ptset, depth}$ procedure in line~6 enumerates all tables of depth $d$ by enumerating operators and tables in $\ptset$.

\newcommand{\matchedrows}{rows}

\smalltitle{Backward Searching}
The backward searching synthesizes the mapping program $\sympm$.
We maintain a set $\pmset = \{ \langle \symsm, h \rangle \} $ for each synthesized mapping statement $\symsm$ and the sub-table $h$ it outputs.
Specifically, we use a structure \emph{HypothesisGenerator} (detailed in \Cref{subsec:optimizations}) to decompose $\texo$ into a prioritized and bounded queue of sub-tables (\emph{hypotheses})~\footnote{In our evaluation, the bound is 20, \ie, we try synthesizing with at most 20 hypotheses for each depth.},
where each element $h \subseteq \texo$ (lines~7--9 of \Cref{alg:top}).
Then, each iteration chooses a hypothesis $h$ and matches it against the intermediate tables in $\ptset$ (line~10).
If a successful mapping is found and a mapping statement $\symsm$ is returned,
we add the synthesized $\symsm$ and hypothesis $h$ to $\pmset$ (lines~10--12) and update the ranking in the hypothesis generator accordingly (line 13).

\newcommand{\symrowmap}{r}
\newcommand{\symcolmap}{c}
\newcommand{\rowmap}[1]{\symrowmap(#1)}
\newcommand{\colmap}[1]{\symcolmap(#1)}
\newcommand{\freevars}{\overline{a}}
\newcommand{\colcombo}{\overline{\col}}
\newcommand{\tpj}{t_{\square}}

\smalltitle{Table Matching}
The forward and backward searching meet in the middle via table matching by
$\Call{Match}{h, \ptset}$ (line~10), which verifies whether a hypothesis $h$ can be produced by projecting some table from $\ptset$.

Specifically, $\Call{Match}{h, \ptset}$ iterates each table $t$ in $\ptset$, and checks whether there exist projection expressions $\pj \ldots$
such that $\eval{\text{\opYield}(t, \pj \ldots)} = h$.
To ease the illustration, we introduce a table $\tpj$ abstracting all columns that can be projected from $t$.
Specifically, $\tpj$ is obtained by appending new columns to $t$, where each new column corresponds to a constant projection or mutate projection.

\[\tpj = t \append \sum_{c} \col_{c} \append \sum_{\highfeature} \sum_{\colcombo} \highfeature(\freevars)(\colcombo) \]

Here, $\sum$ denotes appending a sequence of columns.
The constant $c$ ranges over the values $c$ in $h$ such that there exist a column in $h$ having only one value $c$.
$\col_{c}$ denotes a column filled with constant $c$.
$\highfeature$ ranges over the higher-order features in \Cref{eq:features} and $\freevars$ denotes the \emph{free variables} to parameterize $\highfeature$,
\eg, $regex$ for \codef{substring}.
$\colcombo$ ranges over the column combinations in $t$ that are type consistent with the feature $f = \highfeature(\freevars)$.
Recall that $f(\colcombo)$ denotes the column obtained from \opMutate($f, \colcombo$).

For each higher-order feature $\highfeature$, we introduce free variables $\freevars$ to denote the parameterized feature $\highfeature(\freevars)$
and also the values obtained by evaluating $\highfeature(\freevars)$ on columns $\colcombo$.
For example, consider a column with values $\tuple{2, 3}$, and the higher-order feature is $\texttt{linear}$ that accepts parameters $(a, b)$ and outputs a feature $\lambda x. ax + b$.
If we introduce free variables $a_1$ and $b_1$,
the parameterized feature is denoted by $\lambda x. a_1 * x + b_1$,
and the resulting column is denoted by $\tuple{2a_1+b_1, 3a_1+b_1}$.

With this representation of $\tpj$, we can model the matching between $h$ and $\tpj$ as a constraint-solving problem.
To ease the illustration, we assume the rows and columns in $\tpj$ and $h$ are indexed.
Suppose $\tpj$ is an $m \times n$ table, $h$ is an $m' \times n'$ table ($m \geq m' \land n \geq n'$), %
our target is to find a surjection $\rowmap{i} : [1 \ldots m] \to [1 \ldots m']$ from rows of $\tpj$ to rows of $h$,
and a mapping $\colmap{j} : [1 \ldots n'] \to [1 \ldots n]$ from columns of $h$ to columns of $\tpj$,
such that
\begin{equation}
  \exists \ \symrowmap,\ \symcolmap,\ \paramset. \bigwedge_{
    \scriptsize
    \begin{array}{c}
      1 \le i \le m \\
      1 \le j \le n'\\
    \end{array}
  }
  \tpj[\row_i,\col_{\colmap{j}}] = h[\row_{\rowmap{i}},\col_j],\label{eq:matching}
\end{equation}
where $\paramset$ is the set of free variables.

To solve the constraint, we first enumerate the mappings $\symrowmap$ and $\symcolmap$, then solve the values in $\paramset$.
Specifically, for integral constraints, we solve the integer parameters with an off-the-shelf constraint solver (\eg, Z3~\cite{z3} in our implementation).
For string constraints, we solve the string manipulation program by adopting the algorithm in FlashFill~\cite{flashfill11}.

\smalltitle{Assembling Program}
Once a hypothesis is checked feasible (line~11),
$\Call{AssembleMapping}{\pmset, \texo}$ checks whether the found hypotheses (sub-tables) form a union equal to $\texo$
and outputs the corresponding mapping program $\sympm$ (line~14).
If $\sympm$ is found, then $\Call{AssembleProgram}{\ptset, \sympm}$ generates $\sympt$ from the tables $\sympm$ relies on (select statements from $\ptset$) and returns $\tuple{\sympt, \sympm}$ as the final result (lines~15--16).

\begin{theorem}
(\textbf{Soundness}) Given input-output examples $\tuple{\texi, \texo}$, and $\symprog$ is generated from the algorithm $\text{Synthesize}(\tuple{\texi, \texo})$,
then $\symprog$ is consistent with $\tuple{\texi, \texo}$, \ie, $\eval{\symprog(\texi)} = \texo$.
\end{theorem}

\begin{theorem}
(\textbf{Completeness}) Suppose the bound of hypotheses is unlimited, given input-output examples $\tuple{\texi, \texo}$ such that $\exists \symprog.\ \eval{\symprog(\texi)} = \texo$,
$\text{Synthesize}(\tuple{\texi, \texo})$ must return a program.
In practice, with the limited bound, \synthfunc is not complete, \ie, it may not return a program even if a consistent program exists.
\end{theorem}

\subsection{Optimizations}\label{subsec:optimizations}

\smalltitle{Ranking Heuristics}
The number of hypotheses to search in backward searching is exponential to the number of rows in $\texo$, \ie, the number of elements in the power set of $\texo$.
In our evaluating PBE tasks, the number of rows (user actions) could be up to 30, which makes the number of subsets too massive for brute-force enumeration.
Therefore, we adopt heuristics-based ranking to prioritize the hypotheses to search.

\begin{figure*}[b]
  \begin{center}
    \includegraphics[scale=0.4]{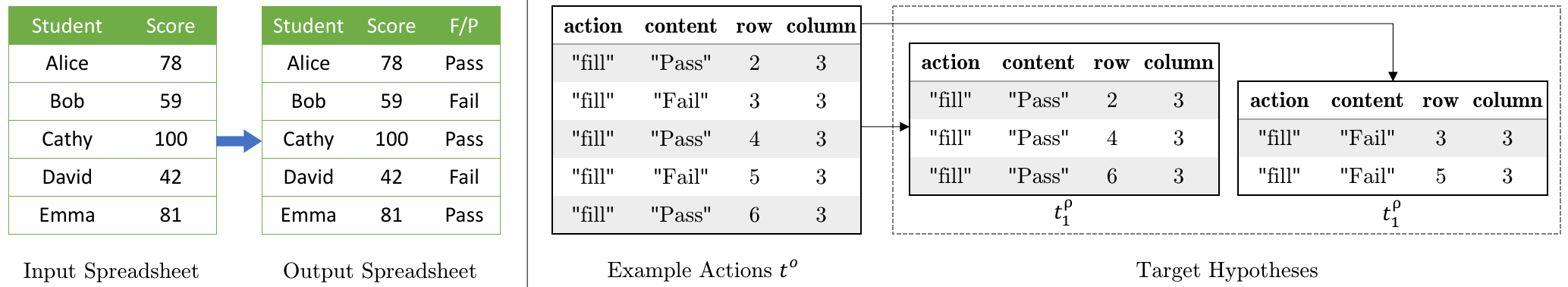}
  \end{center}
  \caption{Example PBE task for illustrating the heuristics. Given a spreadsheet of students and scores, the task needs to classify scores $>= 60$ as ``Pass'' and $< 60$ as ``Fail'' and append the status to the third column.
  The left-hand side shows five examples. $\texo$ denotes the example actions. $\tpja$ and $\tpjb$ denote the target hypotheses that should be ranked high by the heuristics. }
  \label{fig:example-heuristics}
\end{figure*}

To better illustrate the heuristics, we introduce one more PBE task in \Cref{fig:example-heuristics} that aims to categorize Pass/Fail scores.
In the following, we illustrate our heuristics with \Cref{fig:example-heuristics}.

The first heuristic is that we assume the hypotheses generated by different \opYield statements have no intersection.
This assumption usually would not miss the correct programs because, in practice,
user actions produced by different \opYield statements are for processing different data entities.
For example, the two \opYield statements in our motivating task are for processing odd and even frames separately.
Also, the task in \Cref{fig:example-heuristics} processes each spreadsheet row separately.

On the other hand, if the overlap between hypotheses is allowed, the synthesized program is likely to be overfitting.
A typical case of over-fitting is that the program comprises many \opYield statements, each covering a small subset of $\texo$.
When a hypothesis has fewer rows, it is more likely to find a projection that satisfies \Cref{eq:matching}, even though it is not the user's expectation.
A naive case is that when there is only one row in a hypothesis, the corresponding \opYield statement can be simply filled with constant projections.
A program composed of all such \opYield statements is likely to be over-fitting and what we try to circumvent.

Our second heuristic is that if $t$ is one of the target sub-table in $\texo$, then the values in each column of $t$ should be from the same projection.
For example, among the four user actions in \Cref{fig:task-io}, $-30$ and $-40$ are computed from $A_1$, while $+30$ and $+40$ are computed from $A_2$.
Based on this intuition, the hypothesis generator ranks the sub-tables of $\texo$ with the consistent score defined in \Cref{fig:score}.

For each column in a sub-table $t$, we check if it can be a constant column ($S_{const}$) or an order column ($S_{ord}$) or a concatenation column ($S_{concat}$).
By instinct, these conditions check whether the values in the column are from the same projection.
For example, the \codef{content} columns of $\tpja$ and $\tpjb$ are captured by $S_{ord}$.
$S(t)$ is proportional to the number of rows in $t$ because more rows tend to reduce the probability of over-fitness, as we discussed above.
Nonetheless, over-fitness may also happen, \eg, $\texo$ as a whole is scored high because the \codef{row} column happens to be consecutive.

We implemented the above heuristics in a \emph{hypothesis generator} (line~7 in \Cref{alg:top}).
A hypothesis generator takes in the input example $\texi$ and the output table $\texo$, ranks the sub-tables of $\texo$, and iterates them in order of ranking scores.
In addition to the consistency score, we optimize the ranking result using the following measures:
\begin{itemize}
\item If a table $t$ is a subset of a table $t'$ previously iterated and $t'$ has been matched, rank down $t$.
This measure prevents the subsets of a top-ranked table from dominating the top positions
because if $t'$ has high scores in $S_{const}, S_{ord}$, and $S_{concat}$, its subsets also tend to have high scores.
For example, when $\tpja$ has been matched, the two-row subsets of $\tpja$ may also be ranked high but more attention should be paid to tables with other rows, \eg, $\tpjb$.

\item If a table $t$ is matched successfully with a table generated forward,
the generator would rank the complement sub-tables higher (line~13 in \Cref{alg:top}).
This measure is effective if some target sub-tables fail to get high scores.
For example, $\tpjb$ only has two rows and is ranked lower than some wrong sub-tables, \eg, the sub-tables with \codef{row} number ``2-3-4'', ``3-4-5'', and ``4-5-6''.
But when $\tpja$ is matched successfully, we could prioritize the matching of $\tpjb$, which is the complement of $\tpja$.
\end{itemize}

\begin{figure}
  \footnotesize
  \begin{center}
    \begin{equation*}
      \begin{split}
        &S(t) = \sum_{col \in t} [S_{const}(col) + S_{ord}(col) + S_{concat}(col)] * rows(t) \\
        &S_{const}(col) =
        \begin{cases}
          1, & \text{if all values in $col$ are identical}\\
          0, & \text{otherwise}
        \end{cases}\\
        &S_{ord}(col) =
        \begin{cases}
          1, & \text{if the values in $col$ are consecutive integers}\\
          0, & \text{otherwise}
        \end{cases}\\
        &S_{concat}(col) =
        \begin{cases}
          1, & \text{if the strings in $col$ can be projected by a \texttt{concat} feature on some rows in $\texi$}\\
          0, & \text{otherwise}
        \end{cases}
      \end{split}
    \end{equation*}
  \end{center}
  \caption{Equations measuring the consistency score $S(t)$ of table $t$. $rows(t)$ denotes the number of rows in $t$.}
  \label{fig:score}
\end{figure}

\smalltitle{Implementation}
We briefly introduce other optimizations in our implementation here.
First, we noticed that the tables generated in the forward searching have many duplicates because different predicates used in the filter construct may produce the same tables.
In our implementation, we merged predicates and tables into equivalent classes and only considered one in the search process.
We also noticed many same columns could be produced from the \opGroup and \opOrder construct, and we merged the equivalent columns.

Second, the solving of constraints usually contains a lot of common sub-constraints.
We cached the previously computed results to avoid re-invoking the same costly constraint-solving processes.

Third, as some features preserve the order after transformation (\eg, linear, divide, and sum),
we can prune the search by matching and solving the constraints instead of enumerating all the possible permutations.

Fourth, when generating columns from \opOrder, we do not explicitly enumerate all parameters,
we fix $c_{start} = 0$ and $c_{reverse}=false$ to generate columns,
then leverage the feature-solving procedure to synthesize the actual parameters.

%% file: 5_evaluation.tex
\section{Evaluation}\label{sec:evaluation}

In this section, we evaluate the performance and usability of \proj with the following research questions:
\begin{itemize}[leftmargin=*,itemsep=1pt, topsep=3pt]
    \item \textbf{RQ1: (Effectiveness of \proj)} How effective is \proj in solving PBE tasks in multiple domains?
    How is it compared to implementing known DSLs and algorithms on the state-of-the-art PBE frameworks (an alternative for developers to realize PBE for their applications)?
    \item \textbf{RQ2: (Effectiveness of backward searching)} How does the backward searching component contribute to the effectiveness of \proj?%
    \item \textbf{RQ3: (Usability)} Compared with the existing PBE framework, does \proj make PBE more approachable for software developers?
\end{itemize}

To evaluate the effectiveness of \proj, we collected PBE tasks in three different domains as benchmarks and evaluated the success rate of different techniques (\ie,  the proportion of tasks that can be synthesized within a timeout).
To answer RQ1, we compared the PBE tools implemented on top of \proj with those implemented on top of a state-of-the-art PBE framework, \prose~\cite{flashmeta15, flashmeta18neural}.
To answer RQ2, we compared two versions of \proj with and without using the heuristic hypothesis generator.
To answer RQ3, we conducted a human study where participants were asked to implement two PBE tools based on \proj and \prose and then report their assessment of the frameworks' usability in a survey.

\subsection{RQ1 \& RQ2: Effectiveness of \proj}\label{subsec:evaluation-effectiveness}

\subsubsection{Experiment Setup}\label{subsec:implementation}
In this section, we explain our experiment setup to answer RQ1 and RQ2.

\smalltitle{Data Collection}
To evaluate the effectiveness of \proj, we collected 64 benchmarks
in the domains of file management (22), spreadsheet transformation (22), and XML transformation (20).
These three domains were chosen because (1) they represent different forms of data: set, tabular, and hierarchical;
(2) PBE tasks in these domains can be obtained from the literature or online forums.
Each benchmark is a PBE task that contains a task description and input/output examples.
The benchmarks are available on our project website~\cite{beehomepage}.

We collected the benchmarks from existing studies and online forums.
All 22 file management benchmarks are extracted from an earlier study~\cite{file16}.
However, we have no access to the original evaluation dataset, so we recovered a dataset from their task descriptions.
Two tasks were excluded due to their ambiguous descriptions.
As for spreadsheet transformation, seven benchmarks were collected from examples in earlier studies~\cite{table11, flashrelate}.
Similar to file management benchmarks, we recovered the benchmarks from the task descriptions in the papers.
We further augmented the benchmarks with another 15 PBE tasks collected from an online forum, Excel Forum~\cite{excelforum}.
All the 20 XML transformation benchmarks were collected from an online forum OxygenXML~\cite{oxygenxml}.
We collected these benchmarks from the latest posts on the forums when we conducted the experiments.
Posts were selected when they contained a clear description and specific input/output examples.
For each benchmark, we manually inspect the task and specify the values in constant pools.

\smalltitle{Baselines} To answer RQ1 and RQ2, we compared \proj with two baselines, respectively:

\prose: To answer RQ1, we compared \proj with \prose. %
\prose is a state-of-the-art PBE framework developed and maintained by Microsoft. The framework serves as backends for different PBE applications~\cite{merge_conflict21, program-transform17, flashfill11}.
It allows developers to develop PBE tools by providing DSLs and functions to optimize the synthesis algorithms of \prose with domain-specific knowledge.
\prose is a general-purpose framework that supports arbitrary DSL and algorithms, which differs from \proj.
The comparison aims to assess the relative merits of \proj against a simulated scenario of using a SOTA framework (\ie, \prose) to implement known DSLs and algorithms (detailed below).

\projf: To answer RQ3, we introduced \projf, a variant of \proj without the backward searching component,
\ie, the baseline algorithm introduced in \Cref{subsec:algo-motivation}.
All the other optimizations are preserved.

\smalltitle{PBE Tool Implementations}
Both \proj and \prose are frameworks allowing a PBE tool to be constructed with additional customization by developers. The following customization is adopted in our evaluation.

\textit{Customization on \proj.}
As mentioned, the 64 benchmarks collected span across three application domains. We, therefore, prepare a PBE tool for each domain by customizing \proj with the concerned data schema and user actions in tables.
The customization is made using the C\# API interface provided by \proj. %

\textit{Customization on \prose.}
We customize \prose to implement three state-of-the-art DSL-based domain-specific PBE tools as baselines.
For each tool, we provide a grammar file specifying the DSL,
a set of files expressing the semantics of the DSL elements, and callback functions.
Specifically, the DSLs and algorithms for file management and XML transformation were extracted from \hades~\cite{file16}.
The DSL and algorithm for spreadsheet transformation were extracted from a study on synthesizing spreadsheet layout transformation~\cite{table11}.
While other spreadsheet transformation synthesizers~\cite{table-component17, autopandas} exist, they are not completely DSL-based and therefore are not considered.

\smalltitle{Evaluation Metrics}
We evaluated the performance of a synthesizer by measuring:
(a) the success rate: the proportion of benchmarks that can be solved within a timeout (120s), \ie, programs matching the input/output examples were successfully generated, %
(b) the time taken to solve a benchmark, and (c) the number of over-fitting solutions.
A solution to a benchmark is over-fitting if the synthesized program does not capture the intent of the benchmark.
We checked whether a synthesized program is over-fitting by manually comparing the semantics of the program against the description in each benchmark.

All the experiments were conducted on a Windows machine with a 4GHz AMD 5800X CPU and 32GB memory, simulating the ordinary computing power of end-users.
The timeout of 120s is sufficient concerning the response time limit (10s) for keeping the user's attention~\cite{response-time1} and the time limit (60s) for the user to complete simple tasks~\cite{response-time2}.

\subsubsection{Experiment Results}
\begin{figure*}
    \begin{center}
        \begin{subfigure}[b]{0.3\textwidth}
            \centering
            \includegraphics[scale=0.3]{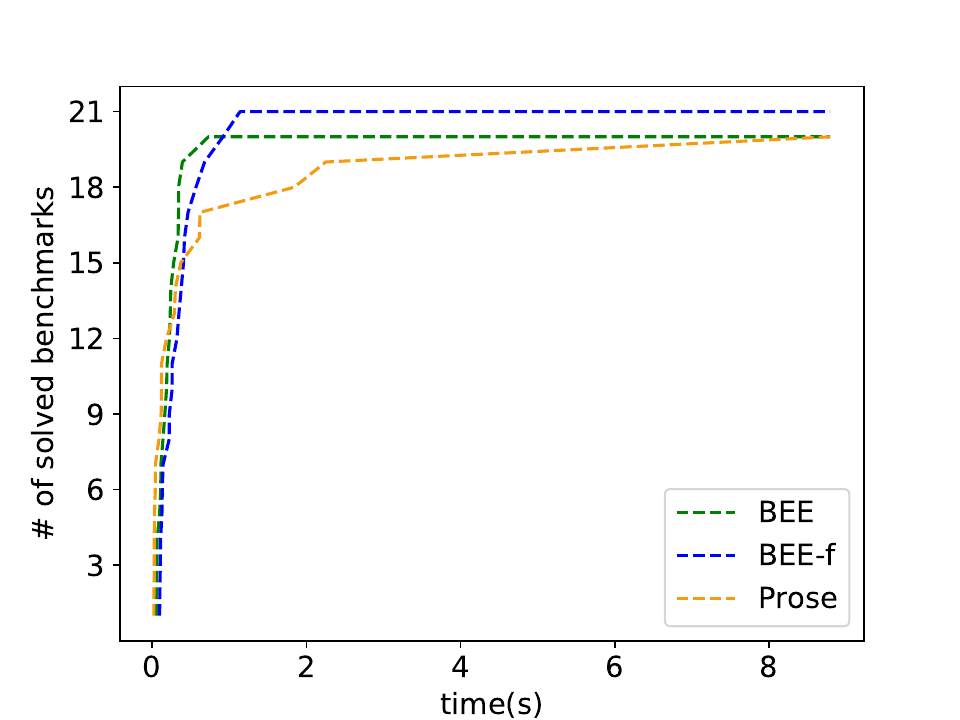}
            \caption{File management.}
            \label{fig:file_result}
        \end{subfigure}
        \hfill
        \begin{subfigure}[b]{0.3\textwidth}
            \centering
            \includegraphics[scale=0.3]{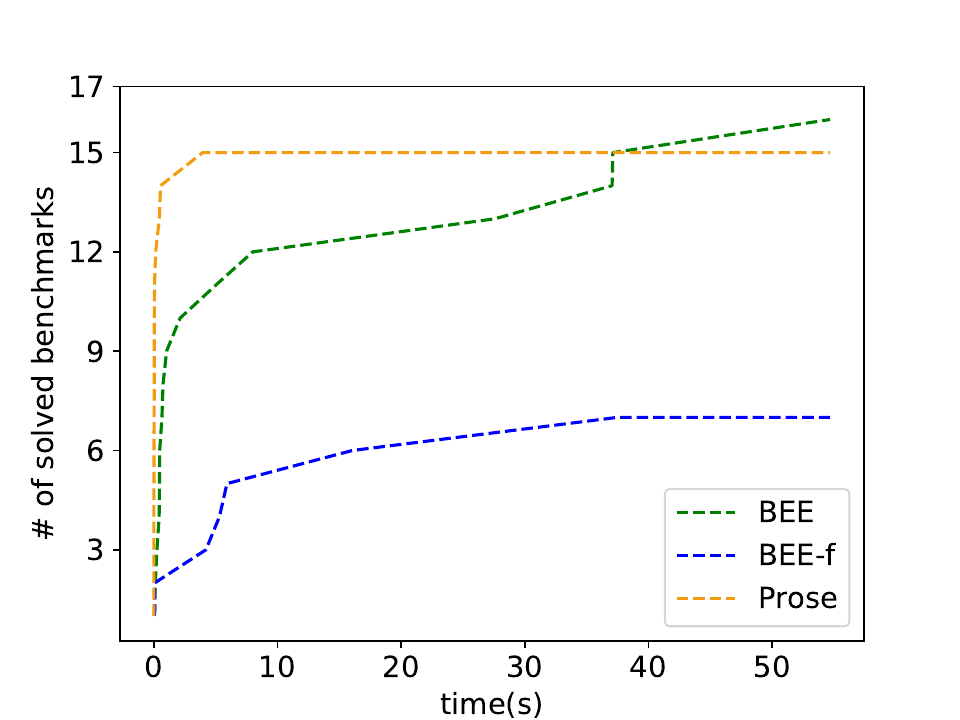}
            \caption{Spreadsheet transformation.}
            \label{fig:spreadsheet_result}
        \end{subfigure}
        \hfill
        \begin{subfigure}[b]{0.3\textwidth}
            \centering
            \includegraphics[scale=0.3]{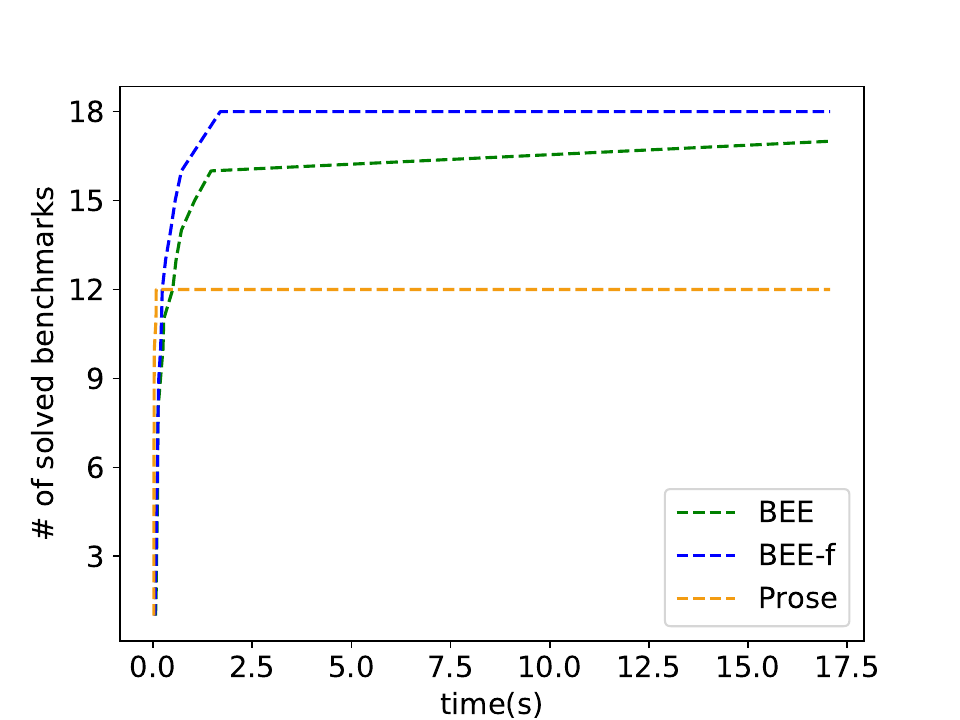}
            \caption{XML transformation.}
            \label{fig:xml_result}
        \end{subfigure}
        \caption{Experiment results on benchmarks under different algorithm settings.
        The $x$ axis indicates the time (seconds) taken to solve the benchmarks.
        The $y$ axis indicates the accumulated number of benchmarks that can be solved.
        The green lines denote \proj's tools, the blue lines denote \projf's tools, and the orange lines denote \prose's tools.
        }
        \label{fig:result}
    \end{center}
\end{figure*}

Table~\ref{tab:success-rate} shows the success rate of \proj and the baseline techniques in the three domains.
Within the timeout (120 seconds), the PBE tools powered by \proj were able to solve 53 out of the total 64 (82.8\%) benchmarks, outperforming that of \prose and \projf (73.4\% and 71.9\%, respectively).
Moreover, all the benchmarks successfully solved by \proj were finished within one minute, and 48 benchmarks (75.0\%) were finished within 10 seconds (Figure~\ref{fig:result}).
Among the successfully solved benchmarks, the longest time taken in file management, spreadsheet transformation, and XML transformation benchmarks is 0.7s, 53.7s, and 15.5s, respectively.

\begin{table*}[]
    \footnotesize
    \caption{Success Rates and Over-fitting Rates of Different PBE Techniques for Benchmarks in Three Domains}
    \label{tab:success-rate}
    \begin{tabular}{@{}lrrrrrr@{}}
        \toprule
        \multicolumn{1}{c}{\multirow{2}{*}{\textbf{Benchmarks}}} & \multicolumn{3}{c}{\textbf{Success Rate}}                                                                                                 & \multicolumn{3}{c}{\textbf{Over-fitting Rate}}                                 \\ \cmidrule(l){2-7}
        \multicolumn{1}{c}{}                            & \multicolumn{1}{c}{\proj} & \multicolumn{1}{c}{\prose} & \multicolumn{1}{c}{\projf} & \multicolumn{1}{c}{\proj} & \multicolumn{1}{c}{\prose} & \multicolumn{1}{c}{\projf} \\ \midrule[0.2pt]
        File management                                            & 20/22 (90.9\%)                           & 20/22 (90.9\%)                            & 21/22 (95.5\%)                            & 1/20(5.0\%)          & 4/20(20.0\%)           & 1/21(4.8\%)        \\
        Spreadsheet transformation                                 & 16/22 (72.7\%)                           & 15/22 (68.2\%)                            & 7/22 (31.8\%)                             & 3/16(25.0\%)         & 7/15(46.7\%)           & 2/7(28.6\%)         \\
        XML transformation                                         & 17/20 (85.0\%)                           & 12/20 (60.0\%)                            & 18/20 (90.0\%)                            & 3/17(17.6\%)         & 0/12(0.0\%)            & 3/18(16.7\%)          \\ \midrule[0.2pt]
        Total                                                      & 53/64 (82.8\%)                           & 47/64 (73.4\%)                            & 46/64 (71.9\%)                            & 7/53(13.2\%)         & 11/47(23.4\%)          & 6/46(13.0\%)         \\ \bottomrule
    \end{tabular}
\end{table*}

We manually inspected all the synthesized programs and found that seven of the synthesized programs over-fitted the provided examples.
\proj over-fitted the benchmarks because they did not provide sufficient input/output examples, \eg, three of them only contained one input/output example.
We further provided examples for these seven benchmarks, and six over-fitting programs were corrected.

\begin{figure}
    \begin{center}
        \includegraphics[width=0.35\linewidth]{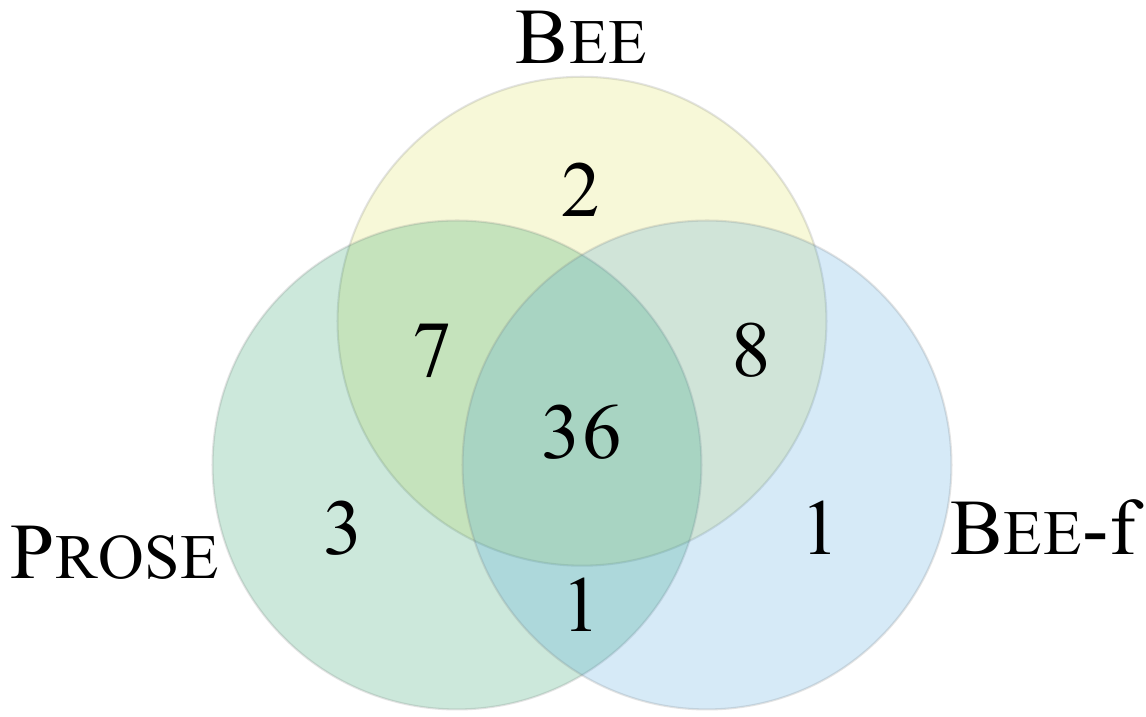}
    \end{center}
    \caption{Venn diagram of benchmarks solved by each technique. For example, 36 benchmarks were solved by both the three techniques.}
    \label{fig:venn}
\end{figure}

We studied the remaining 11 failed benchmarks to inspect the reasons for the failures.
Five of them cannot be well expressed by our current language, and two of them timed out due to their intrinsic complexity.
Typically, the target programs of these benchmarks involve feature computation mixed with complex structural transformations such as join.
The remaining four cases cannot be captured by the optimization heuristics we proposed.
Figure~\ref{fig:venn} shows the Venn Diagram plotting the relationship between benchmarks that can be solved by each technique.
As shown in the figure, among the 11 benchmarks failed by \proj, five of them can be solved by \prose or \projf.
Two benchmarks can be solved by \projf but not \proj.
Our backward searching component failed to decompose appropriate sub-tables for these benchmarks because their output examples were simple and indistinguishable in projection.
This enlightens us to improve the strategy further to generate hypotheses in backward searching by considering more information from the input.
As for \prose, two of the three uniquely solved benchmarks contain operations beyond the expressiveness of \proj,
and the remaining one was too complex to be solved by \proj's synthesis algorithm.
In the future, we plan to further improve the expressiveness of \proj by incorporating more constructs and further optimizing our synthesis algorithm.
In the following, we compare \proj with the two baselines in detail and answer RQ1 and RQ2.

\smalltitle{Comparison with \prose} %
\proj outperformed \prose with regard to both the success rate and the over-fitting rate.
As shown in Table~\ref{tab:success-rate}, overall, \proj achieved a higher success rate (82.8\% v.s. 73.4\%) and a lower over-fitting rate (13.2\% v.s. 23.4\%).
\prose had success rates similar to \proj in the domains of file management and spreadsheet transformation but had more over-fitting cases in both of these domains (4 v.s. 1 and 7 v.s. 3).
\prose had higher over-fitting rates in these domains because the domain-specific algorithms described in their papers did not incorporate ranking strategies~\cite{file16,table11}.
Thus, over-fitting solutions were first synthesized and outputted in these cases.
In the domain of XML transformation, \prose had no over-fitting cases but had a success rate much lower than \proj.
This is because many tasks in the domain of XML transformation are out of the scope of the domain-specific DSL in \prose adapted from \hades~\cite{file16}.
Eight of our XML transformation tasks violate the assumption adopted by the DSL and cannot be expressed.
In comparison, \proj successfully synthesized five of these tasks.
This shows that the expressiveness of \proj is satisfying.
Many real-world XML transformation tasks out of the scope of existing domain-specific DSLs can still be supported by \proj.

Figure~\ref{fig:result} plots the number of solved benchmarks as time passes.
The green line represents \proj, and the orange line represents \prose.
The performance of \proj is similar to that of \prose.
\prose grows faster than \proj only in the spreadsheet transformation domain.
This is because the synthesizer built atop \prose considers a more limited search space optimized for spreadsheet transformation.
\proj cannot leverage such domain-specific optimization since we aim to hide the sophisticated design of DSL and synthesis algorithms by adopting a more inclusive language.
However, the evaluation results show that \proj achieved a good performance comparable to \prose even without domain-specific optimizations.
This shows that the bidirectional synthesis algorithm we proposed helped \proj to synthesize the target programs effectively.

\smalltitle{Comparison with \projf}
As shown in Table~\ref{tab:success-rate}, \proj achieved a higher overall success rate and a similar over-fitting rate compared with \projf.
Specifically, \proj significantly outperformed \projf in the domain of spreadsheet transformation (72.7\% v.s. 31.8\%).
In the domains of file management and XML transformation, \proj had s slightly worse performance than \projf: \projf solved one more benchmark in both domains.
\proj failed to solve these two benchmarks because the output examples in these benchmarks were small-size ($\leq$ six rows in $\texo$)
and indistinguishable.
The backward searching component of \proj failed to decompose the output examples into sub-tables correctly and misled the forward searching.
Instead, the brute-force forward searching adopted by \projf remained manageable due to the relatively small number of output rows.
However, in the domain of spreadsheet transformation, the benchmarks are more complicated, with usually more than ten rows in $\texo$.
\projf timed out for most tasks because the forward searching cannot scale without the guidance of the backward searching.
These results show that our bi-directional searching strategy can improve the effectiveness of the synthesizer, especially for complicated PBE tasks.
To mitigate the issue induced by failures of the backward searching, we can adopt an adaptive strategy that falls back to using only forward searching
for simple tasks with a limited number of output examples.

To summarize, PBE tools implemented on top of \proj achieved better performance than those implemented with the state-of-art PBE framework and integrated domain-specific optimizations.
Our results also show that the bidirectional search strategy adopted by \proj can improve the effectiveness of the synthesis algorithm, especially for complicated tasks.

\subsection{RQ3: Usability}\label{subsec:user-study}
In RQ3, we aim to evaluate whether \proj can ease the process of implementing PBE tools for software developers.
To achieve this, we conducted a human study with 12 participants and compared the usability of \proj
and \prose.

\smalltitle{Participants}
We recruited 12 participants who have four to ten years of coding experience.
They have different levels of familiarity with context-free grammar (CFG, a premise of DSL design in \prose).
Three participants are unfamiliar, six have a basic understanding, and three have used CFG in their projects.
Regarding their SQL familiarity, one is unfamiliar, four have a basic understanding, and seven have used SQL in their projects. %
This confirmed our assumption that SQL is a common skill among programmers.
All the participants have no prior experience in PBE.
They are suitable human study subjects because the target developers of \proj are those with mature programming skills but limited PBE-specific knowledge.
All participants received the same compensation after the human study.

\smalltitle{Methodology}
In our human study, we asked our participants to implement PBE tools with \proj and \prose.
It is impractical for the participants to implement a complete PBE tool in laboratory settings since it can take hours. Confounding factors like participants' level of concentration may arise during the human study. %
To address this problem, we first implemented several PBE tools and removed parts of them to form programming tasks.
The participants only need to finish the programming tasks instead of implementing complete PBE tools.
We prepared PBE tasks as test cases to examine whether a participant's implementation is correct or not.
An implementation is considered correct if it can synthesize programs matching our intended PBE tasks (i.e., passing the PBE test cases).
\newcommand{\tfb}{{F$_B$}\xspace}
\newcommand{\tfp}{{F$_P$}\xspace}
\newcommand{\tsb}{{S$_B$}\xspace}
\newcommand{\tsp}{{S$_P$}\xspace}

Each participant used both \proj and \prose and rated the difficulties in learning and using both frameworks.
A participant used the two frameworks to implement PBE tools for two domains (file management and spreadsheet transformation), respectively.
These two domains were chosen so that the experience in using one framework will not affect the experiment results of using the other.
However, since the difficulty in implementing PBE tools for the two domains can be different, we divided the participants into two groups:
Group~1 used \proj to implement file management PBE tools (\tfb) and \prose to implement spreadsheet transformation (\tsp). Group~2 used \prose to implement file management PBE tools (\tfp) and \proj to implement spreadsheet transformation (\tsb).
We compared the human study results of the two groups to assess the impact of the difficulties of the two domains on participants' performance.
We did not divide the participants into two groups and let each group use and rate only one PBE framework because the criteria to evaluate usability can vary across different participants, and their ratings may not be comparable.

Participants carried out the tasks in different timeslots with no overlap.
Each participant was required to read tutorials of \proj and \prose (with framework names anonymized) before participating in the study to gain a basic understanding of the two frameworks.
The on-site human study started with a discussion with the participant to solve her/his puzzles on the tutorials,
followed by a quiz to ensure the participant has a correct and sufficient understanding before doing the programming tasks.
For each programming task, we introduced the design of the PBE tool, which portions were missed, and the PBE tasks (test cases) it needed to solve.
A participant either finished the PBE tool and passed all the test cases within a timeout of 30 minutes or was considered to fail in the task.
After completing two programming tasks, the participant took a survey to report her/his feelings about \proj and \prose.

The human study was conducted in a university and approved by the university's Human Participants Research Panel.
No personally identifiable information was recorded.

\begin{table}[t]
    \caption{\label{tab:exper-time}Time Spent on Programming Tasks}
    \footnotesize
    \begin{tabular}{c | c c c c c c c}
        \toprule
        Group 1 & P$_1$ & P$_3$ & P$_5$ & P$_7$ & P$_9$ & P$_{11}$  & Avg \\
        \midrule
        \tfb &   8  &  3  &  10  &  20  &  18  &  15  &  12.33  \\
        \tsp &  24  &  8  &  13  &  27  &  22  &  24  &  19.67  \\
        \midrule
        Group 2 & P$_2$ & P$_4$ & P$_6$ & P$_8$ & P$_{10}$ & P$_{12}$ & Avg \\
        \midrule
        \tsb &  25  &  >30  &   7  &  9  &   25  &  13  &  18.17  \\
        \tfp &  30  &   30  &  12  &  5  &  >30  &  13  &  20.00  \\
        \bottomrule
    \end{tabular}
    \begin{tablenotes}
        \item All times are in minutes. Each P$_i$ denotes a participant. Avg shows the average minutes and is calculated by taking numbers > 30 as 30.
    \end{tablenotes}
\end{table}

\begin{table}[t]
    \footnotesize
    \caption{Ranks of Difficulties to Learn/Use Frameworks}
    \label{tab:difficulties}
    \begin{tabular}{c | c c c c c c c c c c c c c}
        \toprule
        Learn & P$_1$ & P$_2$ & P$_3$ & P$_4$ & P$_5$ & P$_6$ & P$_7$ & P$_8$ & P$_9$ & P$_{10}$ & P$_{11}$ & P$_{12}$ &  Avg \\
        \midrule
        \proj  &   2  &    2  &   2  &     4  &    1  &     2 &     2 &     1 &     3 &        3 &        2 &        2 &  2.17   \\
        \prose &   4  &    5  &   3  &     5  &    4  &     4 &     2 &     4 &     3 &        3 &        4 &        4 &  3.75   \\
        \midrule
        Use   & P$_1$ & P$_2$ & P$_3$ & P$_4$ & P$_5$ & P$_{6}$ & P$_7$ & P$_8$ & P$_9$ & P$_{10}$ & P$_{11}$ & P$_{12}$ & Avg \\
        \midrule
        \proj  &   1  &    2  &   2  &     3  &    1  &     2 &     2 &     2 &     2 &        3 &        2 &        2   &  2.00  \\
        \prose &   4  &    3  &   3  &     4  &    3  &     4 &     3 &     3 &     3 &        3 &        4 &        3   &  3.34  \\
        \bottomrule
    \end{tabular}

    \begin{tablenotes}
        \item Ranks are: 1 (very easy), 2 (easy), 3 (median), 4 (difficult) or 5 (very difficult).
        Each P$_i$ denotes a participant.  Avg is the average rating.
    \end{tablenotes}

\end{table}

\smalltitle{Results}
We report the time spent by participants to finish programming tasks in \Cref{tab:exper-time}.
The participants mostly spent less time on the programming tasks with \proj than that of \prose in both groups.
Although the removed portions in both frameworks have similar sizes and complexities (usually a variable, predicate, or a single line of code),
understanding the context in \prose projects may consume more time.
Only P$_4$ and P$_8$ spent more time on \proj than \prose.
However, P$_4$ finished the programming task of \prose (\tfp) in 30 minutes, but the implementation hardcoded a few conditions and over-fitted our test cases.
As for P$_8$, the participant had a good understanding of both frameworks and was able to complete the programming tasks fast,
except that the participant asked about a concept specific to \tsb (instead of \proj) during the human study.
The discussion took four to five minutes.

In the survey, participants were asked to rate the difficulties of learning/using \prose/\proj (\Cref{tab:difficulties}).
Nine out of 12 participants ranked \proj easier to learn than \prose,
and 11 ranked \proj easier to use than \prose.
To investigate whether there is a statistical difference between the difficulty measurements of the two frameworks (\proj v.s. \prose),
we adopt the Wilcoxon signed-rank test~\cite{wilcoxon}.
This non-parametric paired difference test is suitable for paired samples with a small size.
The test result on the samples of difficulties to learn (resp., use) \proj and \prose
shows statistical significance ($p < 0.01$) with the exact $p = 0.007$ (resp., $p = 0.002$).
We received four similar comments like
\textit{``\proj is easier to get started with
while \prose
better suits PBE experts''.}
These results show that \proj is easier to learn and use than \prose for software developers without a PBE background.

We also surveyed which part of \prose the participants deemed the most difficult.
Most participants found the components related to the design of DSL and synthesis algorithms the most difficult: Seven found witness functions (functions to customize synthesis algorithms in \prose) the most difficult to understand, and three identified DSL-related components the most difficult.
Another participant indicated that \prose is difficult to use because it requires much PBE background knowledge.
Our communications with the participants also show supporting evidence of these difficulties.
Before starting the programming tasks, we had a briefing to clarify their puzzles on the tutorials of the two frameworks.
All the participants had spent more time in the briefing on questions for \prose, and
half of them (six) expressed that they could not understand the purpose of witness functions.

In contrast, no participants reported that the database-style interface of \proj caused their confusion.
These observations align with our motivation: (1) the major obstacle for software developers without a PBE background to implement PBE tools is the design of DSLs and synthesis algorithms, and (2) existing PBE frameworks like \prose are still difficult to use for such software developers.
To conclude, although \proj still requires domain-specific customization,
the database-style interface lowers the barrier to PBE tool implementation
compared with the traditional interface that requires DSL and synthesis algorithm.

%% file: 6_ending.tex
\section{Discussion}\label{sec:discussion}
\subsection{Threats to Validity}
\smalltitle{Performance} The performance of \prose in our experiments can be affected by the design of DSLs and the implementation of synthesis algorithms in each domain.
To address the threat, we borrowed the DSLs and synthesis algorithms from representative synthesizers of each domain~\cite{file16, table11}.
Since the artifacts are not publicly available, we re-implemented the DSLs and synthesis algorithms according to the descriptions in the papers.
We also made the tools as compatible with our benchmarks as possible, \eg, by adding language elements as long as they do not require changing the synthesis algorithms.

For \proj, its performance can vary depending on the design of the table schemas.
In our evaluation, the table schemas were designed to suit the general tasks in each domain while not over-fitting the tasks in the benchmarks (e.g., we never hard coded any properties that are shortcuts to the desired ones in specific tasks).
Specifically, one of the authors designed the table schemas for the three evaluating domains, and the other authors validate the design.

    Another threat to performance comparison is that specifications (\eg, languages, input-outputs) to synthesizers on \proj and \prose are different,
    the difficulties of synthesizing programs and being not over-fitted are also different.
    Some tasks can be solved in the \prose specification more easily than in the \proj specification, or sometimes the other way around, as reflected in \Cref{fig:venn}.
    However, the purpose of RQ1 is not to strictly compare two algorithms with the same input specification, but to assess whether \proj as a whole can perform comparatively well against its alternative, \ie, implementing known DSLs and algorithms on \prose.
    To this end, we have tried our best to make the synthesizers on \prose compatible with our benchmarks without changing the synthesis algorithms, which simulate the alternative scenarios.
    We have also made our benchmarks and implementations open-sourced~\cite{beehomepage} for reproduction.

\smalltitle{Usability}
In our usability study, the participants only completed parts of the synthesizer implementation as programming tasks.
For example, both the DSLs of \prose and table schemas of \proj were provided to the users in our study.
We did not ask the participants to implement a complete PBE tool because it is likely to take more time than affordable for in-lab studies.
As such, participants can underestimate the difficulties of implementing synthesizers based on \prose and \proj.
However, we asked the participants to learn the whole process of implementing PBE synthesizers based on both frameworks in our tutorials.
This can help them understand and assess the overall difficulties in PBE synthesizer implementations.

We asked each participant to complete two PBE tools based on each framework respectively in one experiment.
There is a risk that participants may learn some knowledge of PBE implementation from their first task, influencing the second task.
To mitigate the influence, we divided the participants into two groups and altered the order of the frameworks used in our formal experiment.
We did not find substantial differences in the assessments of the two groups of participants.
This is because the approaches to implementing PBE tools on top of \prose and \proj are divergent.
Besides, the two tasks they need to solve fall into two different domains.
As a result, the knowledge learned from one framework had a minor impact on the other one.

    \subsection{Expressiveness}\label{subsec:expressiveness}
    With the design goal to maximize the ease of developer interface, \proj adopts the scheme of a predefined meta DSL with customizable entities/actions.
    Consequently, \proj is not as generally applicable as other frameworks that allow DSL customization (\eg, \prose).
    The expressiveness of the meta DSL (\ie, \lang) is limited purposefully to make the synthesis procedure tractable. This also makes \proj inapplicable to certain tasks.
    Although formally describing a language's expressiveness is difficult~\cite{libkin_expressive_sql_2003},
    we could discuss the expressiveness of \lang by comparing it with its basis, a commonly used SQL subset (\ie, select, where, join, group, order) that has been studied~\cite{libkin_expressive_sql_2003} extensively.

    A fundamental limitation in expressiveness inherited by \lang is recursive query, which queries results from itself as a recursive function calls itself.
    For example, given a set of edges $E$ in a graph, query the reachability closure $R$, \ie, node pairs $(a, b)$ such that there is a path from $a$ to $b$. The recursive query can be $(a, b) \in R$ if and only if $(a, b) \in E$ or there exists $c$ such that $(a, c) \in E$ and $(c, b) \in R$ .
    Despite recursive queries that require an unbounded number of nested queries, some tasks requiring a large (but bounded) number of queries are also difficult for \proj to synthesize in practice, although possible in theory.
    For example, given a code repository's abstract syntax trees (ASTs), we could represent the hierarchical structure by introducing a \codef{parent} field to each AST node.
    However, to extract AST subtrees that match a specific pattern,
    the target \lang program may need many joining steps to recover a subtree and compare it with the target pattern.
    Synthesizing such programs is likely to fail due to the overwhelming search space.

    The major difference brought by \proj is the adoption of features in the \opYield statements (\ie, the mapping program $\sympm$).
    This largely strengthens the expressiveness of \lang compared with the base SQL subset, particularly when handling PBE tasks, which generally require computations such as string manipulations.
    Before $\sympm$, the tables that could be expressed/queried in the transformation program $\sympt$ are nearly identical to those queried by the base SQL subset.
    Then, in $\sympm$, we can perform \emph{row-wise} computations to query columns (or called features) that cannot be expressed in the base SQL subset.
    Such computations would not change the original table structure, \ie, no rows would be added/removed, but only new columns would be added.
    The expressiveness of these row-wise computations is defined by the set of features, specifically, integral arithmetics and string manipulations in the current version.
    In a nutshell, the expressiveness of \lang could be viewed as the common SQL queries followed by row-wise feature computations.

    \lang is restrained but can also automate a wide range of repetitive tasks.
    Many existing PBE tools~\cite{visual19, extract14, extract16, extract17, edit19, table11, file16} share de facto similar expressive power with different predicates/features.
    Besides, the interface of relational tables for capturing entities/actions is general and compatible with any language that can generate actions from entities, which allows us to improve the expressiveness of \lang in the future without affecting extant tools built on \proj.

    \subsection{Future Work}
    To enhance the expressiveness of \proj and let it support more PBE tasks, the underlying search space at the synthesis stage could be even larger.
    While the current synthesis algorithm works well for simple cases, it may not scale to complex ones, \eg, those involving four or more nested filter/join/group operators.
    We plan to further improve the algorithm with techniques such as abstraction~\cite{conflict18, learning_abstraction}, search prioritization~\cite{phog}, and data-driven library learning~\cite{ellis_dreamcoder_2021, top_down_library_learning} to improve the performance of \proj.
    Another direction is to leverage interactive synthesis to alleviate the synthesis pressure with additional end-user inputs, \eg, hints on the entity fields related to the task.
    The number of interactions could be minimized by adapting existing techniques~\cite{ji_question_2020, shen_active_2021, shen_using_2019}.

Besides, the inclusive data-action model in \proj has the potential to support cross-domain PBE tasks (\eg, finding intersected items between a webpage and a spreadsheet).
In the future, we will explore leveraging \proj to synthesize PBE tasks that involve data and actions in multiple domains.

\section{Related Work}\label{sec:related-work}

\smalltitle{Domain-specific PBE Techniques}
In the recent decade, PBE has gained increasing popularity and has found applications in various domains
such as spreadsheet~\cite{flashfill11, table11, spreadsheet-types},
information extraction~\cite{extract14, extract16, extract17},
data visualization~\cite{visual19}, version control~\cite{merge_conflict21}, text/code edition~\cite{edit19, program-transform20}, etc~\cite{parser15, drawing14}.
Typically, to make the search space of the programs to synthesize tractable,
each application focuses on a DSL (defined by the researchers) capturing the specific needs of each domain.
Domain-specific techniques often adopt synthesis algorithms tailored for the DSLs to optimize the synthesis performance.
Take the classic string manipulation DSL~\cite{flashfill11} for example.
To synthesize $f$ for $f(\text{``foo'', ``bar''}) = \text{``FooBar''}$, the synthesis algorithm searches for solutions to sub-problems,
\eg, $f_1(\text{``foo'', ``bar''}) = \text{``Foo''}$, $f_2(\text{``foo'', ``bar''}) = \text{``Bar''}$, \etc.
The sub-solutions are then composed into a complete solution, \eg, $f = \lambda x. \codef{concat}(f_1(x), f_2(x))$.
This divide-and-conquer approach is tailored for the commonly used \codef{concat} operator in the string manipulating DSL and may not apply to other DSLs.
Likewise, specialized DSLs and synthesis techniques are generally hard to be transferred from an existing domain to another one.

\smalltitle{Customizable PBE Frameworks}
Several program synthesis frameworks~\cite{flashmeta15, trinity, rosette} have been proposed to facilitate the development of program synthesis tools.
They provide facilities (\eg, APIs) to help developers define a DSL (syntax and semantics) and ship a default synthesis algorithm.
These frameworks are designed for experts with sufficient background knowledge on program synthesis.
For example, to use \rosette~\cite{rosette}, the user needs to understand what operators have support in SMT background theories~\cite{smt-lib2} and design a compliant DSL because the synthesis algorithm in \rosette is based on SMT solving.
To use \prose~\cite{flashmeta15}, the user needs to understand its synthesis algorithm and design a compliant DSL where each production is a function call, e.g., $A \to f(B, C)$.
Usually, the synthesis algorithm also requires the user to provide callback functions, \ie, functions invoked by the synthesis engine to decompose constraints on $f(B, C)$ into constraints on $B$ and $C$.
Tuning the DSL or synthesis algorithm is arduous for non-PBE-experts.
In comparison, \proj provides a developer-friendly interface to make PBE more approachable for PBE novices.
Our evaluation also shows that \proj is easier to learn and use than existing PBE frameworks.

\smalltitle{Table-based Program Synthesis}
There are many synthesis techniques based on table-like data structures,
including SQL query synthesis by examples~\cite{sql09, sql13, sql17, analytical_sql_22} and by natural language~\cite{dong2018coarse, ni2020merging},
table schema refactoring~\cite{sql-refactor}, database access model synthesis~\cite{shen_using_2019, shen_active_2021}, Datalog-like program synthesis~\cite{datalog18, provenance_guided_19, example_guided_21}, and other transformations on tables beyond SQL~\cite{table-component17, conflict18, autopandas, visual19}.
These techniques cannot synthesize complex expressions (\eg, with constants) as we discussed in \Cref{sec:algorithm}.
There are several recent works that aim to synthesize such complex expressions~\cite{data_completion_17, analytical_sql_22, table_graph_21}.
They all require extra specifications inputted from users, \eg, sketches of expressions or demonstration of the computation procedure instead of only the final value.
Such specifications can only be provided by spreadsheet/database users who have a good understanding of the desired programs.
We cannot introduce such specifications to \proj because the majority of applications (\eg, file manager, E-mail client) that \proj may support are facing end-users without a programming background.
We design a bidirectional algorithm for a simplified setting where the expression only exists in the last step.
This design does not require additional inputs from end users and achieved good performance in practice, as demonstrated in our evaluation.

\smalltitle{Synthesis Technique}
The synthesis algorithm of \proj leverages bidirectional search and divide-and-conquer.
Bidirectional search is an effective approach to exploring the program space (\eg, programs defined in a DSL) with respect to the given specification (\eg, examples).
Generally speaking, given a specification in the form of examples $(\xi_{in} = \{i_1, \ldots\, i_n\}, \xi_{out} = \{o_1, \ldots, o_n\})$, to find $P$ such that $P(i_{j}) = o_{j}$ for each $j$,
bidirectional search
    (1) explores the programs and states that derive from $\xi_{in}$ (forward searching),
    (2) explores the programs and specifications that derive from $\xi_{out}$ (backward searching),
    and (3) composes $P$ using specifications explored from the two directions.
Forward searching usually exploits syntax-guided enumeration of the DSL to produce concrete programs and running states,
while the form of specification in backward searching varies from application to application.
For example, \duet~\cite{lee_combining_2021} targets SyGuS~\cite{sygus} programs and recursively deduces the output specification (initially $\xi_{out}$) of functions $f(x_1, \ldots)$ to produce specifications for sub-expressions ($x_1, \ldots$).
\viser~\cite{visual19} targets table transformation programs and leverages table-inclusion constraints (tables explored forward need to subsume the tables explored backward) to prune search space.
Different from them, \proj searches for concrete sub-tables of the target table in the backward searching to support synthesizing the complex value computations that are required in many \proj tasks.
The backward searching process (\ie, searching sub-tables) in \proj follows a divide-and-conquer pattern,
which was also leveraged by the STUN (synthesis through unification) framework~\cite{unification15, divide-conquer} to solve the problem of synthesizing nested \codef{if-then-else} expressions.
STUN synthesizes programs in a bottom-up manner, which first syntactically enumerates expressions for each atomic branch until the expressions cover all input-output examples and finally unifies the expressions using decision-tree learning.
In contrast, \proj decomposes sub-tables in a top-down manner and matches with tables generated forward to synthesize the \opYield statement.
As pointed out by a recent work~\cite{occam_learning_oopsla21}, STUN is likely to synthesize a program by unifying many branches, each covering a small number of examples.
If the constraints of the desired program are given by examples, such programs are likely to be over-fitting, as we analyzed in \Cref{subsec:optimizations}.
In contrast, the ranking strategies of \proj prefer larger sub-tables and thus bias not to synthesize such over-fitting programs.

\section{Conclusion}\label{sec:conclusion}
This paper introduces \proj, a PBE framework making PBE approachable for software developers without PBE expertise.
We formulated \proj's interface, where relational tables serve as a unified representation to model PBE tasks, and developers can easily build PBE tools by customizing the table schemas specific to their needs.
We introduced a DSL \lang that combines table transformation and non-trivial value computation to express PBE tasks.
We designed a bidirectional synthesis algorithm that effectively synthesizes \lang programs.
Our evaluation in three different domains showed that \proj could achieve good performance comparable to that of domain-specific synthesizers.
Moreover, our human study revealed that, compared to a state-of-the-art PBE framework, \proj is easier to use and learn for non-PBE-expert developers.
In the future, we will further improve \proj by enhancing the expressiveness of the language and the performance of the synthesis algorithm.

%% file: 7_ack.tex
\begin{acks}
    \begin{updated}
        We thank the editors and the anonymous reviewers for their constructive comments and suggestions.
        We also thank the participants in the human study and undergraduate students who helped explore the applications of \proj.
        
        This project is supported by the National Science Foundation of China (Nos. 61932021 and 62272218),
        the Leading-edge Technology Program of the Jiangsu Natural Science Foundation under Grant (No. BK20202001),
        the Hong Kong Research Grant Council/General Research Fund (Grant No. 16207120),
        the Hong Kong Research Grant Council/Research Impact Fund (Grant No. R503418),
        and the Natural Sciences and Engineering Research Council of Canada Discovery Grant (Grant Nos. RGPIN-2022-03744 and DGECR-2022-00378).

        This project is also supported by the Fundamental Research Funds for the Central Universities of China (020214912220).
        The authors would like to express their gratitude for the support from the Xiaomi Foundation and the Collaborative Innovation Center of Novel Software Technology and Industrialization, Jiangsu, China.
    \end{updated}
\end{acks}

%% file: a1_example.tex
\section{Running Example Specification}\label{sec:running-example-specification}

In this section, we detail the specifications of the running example in \Cref{fig:task-io} in case readers find some parts of the illustration unclear due to the simplification.
To clarify, the GUI and the entity/action schemas in this example are hypothesized for presenting the idea of \proj and not implemented.

\smalltitle{Entities}
In this task, each frame is an entity.
A frame could be modeled by many fields such as a unique identifier $id$, the order in the frame sequence $frame$, the source file of the figure $file$, the width/length of the figure, etc.
To make the illustration simple, we only consider the following schema with three fields.

\[\symschema = \langle id: \typeid,\ frame: \typeint,\ file: \typestr \rangle \]

The schema can be specified with C\# annotation APIs, as shown in \Cref{lst:data-entities}.

\smalltitle{Actions}
In this task, we only consider a shift action with the following schema.
\[\symschemao = \langle action: \typestr,\ id: \typeid,\ channel: \typestr,\ bx: \typeint,\ by: \typeint \rangle \]
Specifically, the $action$ must be ``shift'' for \proj to recognize.
Executing an action would move the $channel$s (specified by strings such as ``GB'' and ``RGB'') of a frame (specified by $id$) $bx$ pixels right (or left if $bx$ is negative) and $by$ pixels down (or up if $by$ is negative).
The schema can be specified with C\# annotation APIs, as shown in \Cref{lst:user-actions}.

\smalltitle{Input-Output Examples}
The input examples are the first four frames.
They can be modeled by $\symschema$ as the following table $t^i$, and $\texi = \{ t^i \}$.

\begin{center}
    \footnotesize
    \begin{tabular}{| c | c | c |}
        \hline
        \textbf{file} & \textbf{frame} & \textbf{id} \\
        \hline
        ``tiktok.jpg'' & 1 & f1\\
        ``tiktok.jpg'' & 2 & f2\\
        ``tiktok.jpg'' & 3 & f3\\
        ``tiktok.jpg'' & 4 & f4\\
        \hline
    \end{tabular}
\end{center}

The output examples are the four shift actions applied on the first four frames.
The example output table $\texo$, modeled by $\symschemao$, is as follows.

\begin{center}
    \footnotesize
    \begin{tabular}{| c | c | c | c | c|}
        \hline
        \textbf{action} & \textbf{id} & \textbf{channel} & \textbf{bx} & \textbf{by} \\
        \hline
        ``shift'' & f1 & ``GB'' & -30 & -30\\
        ``shift'' & f2 & ``GB'' & +30 & +30\\
        ``shift'' & f3 & ``GB'' & -40 & -40\\
        ``shift'' & f4 & ``GB'' & +40 & +40\\
        \hline
    \end{tabular}
\end{center}

\smalltitle{Target Program}
The target \lang program for handling the running example is as follows.

\lstset{basicstyle=\footnotesize\ttfamily}
\begin{lstlisting}[label={lst:target-prog}]
u = Filter(ti, isOdd(frame));
v = Filter(ti, isEven(frame));

Yield("shift", u, id, "GB", linear(-5, -25)(frame), linear(-5, -25)(frame));
Yield("shift", v, id, "GB", linear(5, 20)(frame), linear(5, 20)(frame));
\end{lstlisting}

Before executing the program, a state $\symstate$ to store table variables is initialized with $\{ti\}$, where $ti$ is bounded to $t^i$.
When executing the first statement \lstinline{u = Filter(ti, isOdd(frame));}, a table $u$ as follows is produced by filtering rows whose frames are odd numbers.
$\symstate$ is updated as $\{ti, u \}$.

\begin{center}
    \footnotesize
    \begin{tabular}{| c | c | c |}
        \hline
        \textbf{file} & \textbf{frame} & \textbf{id} \\
        \hline
        ``tiktok.jpg'' & 1 & f1\\
        ``tiktok.jpg'' & 3 & f3\\
        \hline
    \end{tabular}
\end{center}

When executing the second statement \lstinline{u = Filter(ti, isEven(frame));}, a table $v$ as follows is produced by filtering rows whose frames are even numbers.
$\symstate$ is updated as $\{ti, u, v \}$.

\begin{center}
    \footnotesize
    \begin{tabular}{| c | c | c |}
        \hline
        \textbf{file} & \textbf{frame} & \textbf{id} \\
        \hline
        ``tiktok.jpg'' & 2 & f2\\
        ``tiktok.jpg'' & 4 & f4\\
        \hline
    \end{tabular}
\end{center}

Finally, when executing the last two \codef{Yield} statements, the following two tables $\tpja$ and $\tpjb$ would be generated as sub-tables of the final program output.

\begin{center}
    \footnotesize
    \begin{tabular}{| c | c | c | c | c|}
        \hline
        \textbf{action} & \textbf{id} & \textbf{channel} & \textbf{bx} & \textbf{by} \\
        \hline
        ``shift'' & f1 & ``GB'' & -30 & -30\\
        ``shift'' & f3 & ``GB'' & -40 & -40\\
        \hline
    \end{tabular}
    \quad
    \begin{tabular}{| c | c | c | c | c|}
        \hline
        \textbf{action} & \textbf{id} & \textbf{channel} & \textbf{bx} & \textbf{by} \\
        \hline
        ``shift'' & f2 & ``GB'' & +30 & +30\\
        ``shift'' & f4 & ``GB'' & +40 & +40\\
        \hline
    \end{tabular}
\end{center}

The final output table is the union of tables generated from \codef{Yield}, \eg, the above two tables.

\begin{center}
    \footnotesize
    \begin{tabular}{| c | c | c | c | c|}
        \hline
        \textbf{action} & \textbf{id} & \textbf{channel} & \textbf{bx} & \textbf{by} \\
        \hline
        ``shift'' & f1 & ``GB'' & -30 & -30\\
        ``shift'' & f2 & ``GB'' & +30 & +30\\
        ``shift'' & f3 & ``GB'' & -40 & -40\\
        ``shift'' & f4 & ``GB'' & +40 & +40\\
        \hline
    \end{tabular}
\end{center}

\smalltitle{Synthesis Procedure}
The target program is synthesized in a bidirectional way.

In the forward direction, transform statements are enumerated according to the syntax in \Cref{fig:syntax} with the depth starting from $1$.
The first two \codef{Filter} statements (and the corresponding tables $u$ and $v$) are enumerated when the depth is $1$.
During the synthesis, a set $\ptset$ is used to maintain the enumerated statements/tables, \eg, $\{ \tuple{\epsilon, ti}, \tuple{\codef{Filter(...)}, u}, \tuple{\codef{Filter(...)}, v}, \ldots \}$.

In the backward direction, the expected output table is decomposed into sub-tables using the hypothesis generator introduced in \Cref{sec:algorithm}.
In this running example, the generator ranks and generates top scored tables including the two sub-tables $\tpja$ and $\tpjb$.
Each sub-table corresponds to a \codef{Yield} statement and is matched against tables in $\ptset$.

We illustrate the matching process considering two tables $u$ and $v$ in the forward direction and two tables $\tpja$ and $\tpjb$ in the backward direction.
Before matching $u$ against $\tpja$, we first prepare an abstract table $\tpj$ by adding to $u$ constant columns in $\tpja$ and parameterized features, which is shown as follows.
$a$ and $b$ are integer variables to be resolved.

\begin{center}
    \footnotesize
    \begin{tabular}{| c | c | c | c | c | c |}
        \hline
        \textbf{file} & \textbf{frame} & \textbf{id}  & \textbf{c1} & \textbf{c2} & \textbf{linear} \\
        \hline
        ``tiktok.jpg'' & 1 & f1 & ``shift'' & ``GB'' & $a + b$ \\
        ``tiktok.jpg'' & 3 & f3 & ``shift'' & ``GB'' & $3a + b$ \\
        \hline
    \end{tabular}
\end{center}

Then, we can enumerate the mappings from rows in $\tpj$ to rows in $\tpja$ and mappings from columns in $\tpja$ to columns in $\tpj$.
The correct mappings can be viewed as follows.

\begin{center}
    \footnotesize
    \begin{tabular}{| c | c | c | c | c | c |}
        \hline
        \textbf{c1} & \textbf{id}  & \textbf{c2} & \textbf{linear} & \textbf{linear} & \ldots \\
        \hline
        ``shift''  & f1 & ``GB'' & $a + b$ & $a + b$ & \ldots \\
        ``shift''  & f3 & ``GB'' & $3a + b$ & $3a + b$ & \ldots \\
        \hline
    \end{tabular}
    \quad
    $\to$
    \quad
    \begin{tabular}{| c | c | c | c | c|}
        \hline
        \textbf{action} & \textbf{id} & \textbf{channel} & \textbf{bx} & \textbf{by} \\
        \hline
        ``shift'' & f1 & ``GB'' & -30 & -30\\
        ``shift'' & f3 & ``GB'' & -40 & -40\\
        \hline
    \end{tabular}
    \quad
    \cmark
    \quad
\end{center}

With the mapping, we try solving \Cref{eq:matching} by leveraging off-the-shelf solvers, \eg, SMT solver in this case to solve constraints $a + b = -30 \land 3a + b = -40$.
Finally, we found the correct mappings between rows/columns and the parameter values ${ a = -5 \land b = -25 }$.
The corresponding \codef{Yield} statement can be generated accordingly.

The trial of matching $u$ against $\tpjb$ fails because we cannot find any \typeid-type value in $u$ that can be mapped to f2 or f4 in $\tpjb$.

\begin{center}
    \footnotesize
    \begin{tabular}{| c | c | c |}
        \hline
        \textbf{file} & \textbf{frame} & \textbf{id} \\
        \hline
        ``tiktok.jpg'' & 1 & f1\\
        ``tiktok.jpg'' & 3 & f3\\
        \hline
    \end{tabular}
    \quad
    $\to$
    \quad
    \begin{tabular}{| c | c | c | c | c|}
        \hline
        \textbf{action} & \textbf{id} & \textbf{channel} & \textbf{bx} & \textbf{by} \\
        \hline
        ``shift'' & f2 & ``GB'' & +30 & +30\\
        ``shift'' & f4 & ``GB'' & +40 & +40\\
        \hline
    \end{tabular}
    \quad
    \xmark
    \quad
\end{center}

Similarly, we can find a mapping statement that can map from $v$ to $\tpjb$ but no statement from $u$ to $\tpjb$.
Since we have found successful mappings from $\ptset$ to $\tpja$ and $\tpjb$, and the union of $\tpja$ and $\tpjb$ is exactly the expected output,
we can assemble the program accordingly.

%% file: a2_evaluating_specs.tex
\section{PBE Tool Implementations}\label{sec:pbe-tool-implementations}
We briefly introduce how the PBE tools in the three evaluating domains are implemented on \proj in this section.
Each subsection corresponds to one domain.
For each domain, we introduce the table schemas used to model the data entities and the user actions.
Readers may find full details on the homepage of \proj~\cite{beehomepage}.

\subsection{File Management}\label{subsec:file-management}
Each file entity is modeled as a row in the table.
The fields are listed as follows.

\begin{center}
    \footnotesize
    \begin{tabular}{|c|c|}
        \hline
        \textbf{Field}  & \textbf{Type} \\
        \hline
        id & \typeid \\
        basename & \typestr \\
        extension & \typestr  \\
        path & \typestr  \\
        size & \typeint \\
        modification\_time & Time \\
        readable & Boolean \\
        writable & Boolean \\
        executable & Boolean \\
        group & \typestr \\
        year & \typeint \\
        month & \typeint \\
        day & \typeint \\
        year\_s & \typestr \\
        month\_s & \typestr \\
        day\_s & \typestr \\
        \hline
    \end{tabular}
\end{center}

Different actions have different schemas (arguments).
The actions and arguments they take are listed as follows.

\begin{center}
    \footnotesize
    \begin{tabular}{|c|l|}
        \hline
        \textbf{Action}  & \textbf{Arguments}  \\
        \hline
        chmod & id: \typeid, mod: \typestr \\
        copy & id: \typeid, path: \typestr \\
        unzip & id: \typeid, path: \typestr \\
        move & id: \typeid, path: \typestr \\
        rename & id: \typeid, name: \typestr \\
        delete & id: \typeid \\
        chgrp & id: \typeid, group: \typestr  \\
        chext & id: \typeid, extension: \typestr  \\
        tar & id: \typeid, name: \typestr \\
        \hline
    \end{tabular}
\end{center}

\subsection{Spreadsheet}\label{subsec:spreadsheet}
Given a spreadsheet file and a consecutive range of cells, we model the cells with two tables of different schemas.
The first table is obtained by modeling each cell as a row with the following fields.
The type of row\_head, col\_head, content depends on the type of spreadsheet cell.
The read\_ord field is the reading order of cells (left-right and up-down).

\smallskip
\begin{center}
    \footnotesize
    \begin{tabular}{|c|c|}
        \hline
        \textbf{Field} & \textbf{Type} \\
        \hline
        id & \typeid  \\
        row & \typeint \\
        col & \typeint \\
        row\_head & \typestr | \typeint  \\
        col\_head & \typestr | \typeint  \\
        content & \typestr | \typeint  \\
        read\_ord & \typeint \\
        \hline
    \end{tabular}
\end{center}
\smallskip

The second table is close to the original tabular representation of a spreadsheet range.
As shown in the schema below, each row of the cells corresponds to a row in the output table,
and each column of the cells corresponds to a column in the output table.
The type of value in column col$i$ depends on the type of the $i$-th column of spreadsheet.

\smallskip
\begin{center}
    \footnotesize
    \begin{tabular}{|c|c|}
        \hline
        \textbf{Field} & \textbf{Type} \\
        \hline
        row & \typeint  \\
        col1 & \typestr | \typeint \\
        col2 & \typestr | \typeint \\
        \multicolumn{2}{|c|}{$\dots$} \\
        \hline
    \end{tabular}
\end{center}
\smallskip

The only user action is to fill a cell (given by row and column) with content.
Deleting a cell is equivalent to filling it with empty content.

\smallskip
\begin{center}
    \footnotesize
    \begin{tabular}{|c|l|}
        \hline
        \textbf{Action} & \textbf{Arguments} \\
        \hline
        fill   & content: \typestr, row: \typeint, col: \typeint \\
        \hline
    \end{tabular}
\end{center}
\smallskip

\subsection{XML}\label{subsec:xml}
Given an XML file and a list of selected elements, we represent them with two SQL tables using the schemas below, one for XML elements and one for XML attributes.
Each XML element is modeled as a row in the former table with its id, its tag, its text,
and the ids of its parent, its previous sibling, and its next sibling.~\footnote{If there is no parent/previous/next element, replace with a special null id.}
Each XML attribute is modeled a row in the latter table with its id, the XML element it belongs to, its key, and its value.

\smallskip
\begin{center}
    \footnotesize
    \begin{tabular}{|c|c|}
        \hline
        \textbf{Field} & \textbf{Type}\\
        \hline
        id & \typeid \\
        tag & \typestr  \\
        text & \typestr \\
        parent & \typeid   \\
        previous & \typeid \\
        next & \typeid \\
        \hline
    \end{tabular}
    \quad
    \begin{tabular}{|c|c|}
        \hline
        \textbf{Field} & \textbf{Type}\\
        \hline
        id & \typeid \\
        element & \typeid  \\
        key & \typestr \\
        value & \typestr \\
        \hline
    \end{tabular}
\end{center}
\smallskip

The actions used are listed as follows.
These actions generally follow APIs in manipulating XML elements and attributes.
However, when used in the context of XML editing, end-users may need to provide additional information, \eg, which action does the editing correspond to.

\smallskip
\begin{center}
    \footnotesize
    \begin{tabular}{|c|l|}
        \hline
        \textbf{Action} & \textbf{Arguments} \\
        \hline
        delete\_element & element: \typeid  \\
        modify\_text  & element: \typeid, text: \typestr \\
        modify\_attribute & element: \typeid, value: \typestr \\
        modify\_tag   & element: \typeid, tag: \typestr \\
        add\_element & parent: \typeid, tag: \typestr, text: \typestr \\
        add\_element\_above & element: \typeid, tag: \typestr, text: \typestr \\
        add\_attribute & element: \typeid, key: \typestr, value: \typestr \\
        wrap & element: \typeid, tag: \typestr \\
        move\_below & element: \typeid, target: \typeid \\
        append\_child & element: \typeid, target: \typeid \\
        \hline
    \end{tabular}
\end{center}